\documentclass[a4paper,fleqn,usenatbib]{mnras}

\usepackage{newtxtext,newtxmath}

\usepackage[T1]{fontenc}
\usepackage{ae,aecompl}

\usepackage{graphicx}	
\usepackage{amsmath}	
\usepackage{amssymb}	

\usepackage[flushleft]{threeparttable}
\usepackage[pdftex,usenames,dvipsnames]{xcolor}
\usepackage{natbib}
\usepackage{aas_macros}
\usepackage{url}
\usepackage[none]{hyphenat}
\usepackage{times}

\title[Intermediate-mass black holes in dwarf galaxies out to redshift $\sim$ 2.4]{Intermediate-mass black holes in dwarf galaxies out to redshift $\sim$ 2.4 in the \textit{Chandra} COSMOS Legacy Survey}

\author[Mezcua et al.]{
M. Mezcua,$^{1,2}$\thanks{E-mail: marmezcua.astro@gmail.com}
F.~Civano,$^{3}$
S.~Marchesi,$^{4}$
H.~Suh,$^{5,6}$
G.~Fabbiano,$^{3}$
and M.~Volonteri$^{7}$
\\
$^{1}$Institute of Space Sciences (ICE, CSIC), Campus UAB, Carrer de Magrans, 08193 Barcelona, Spain\\
$^{2}$Institut d'Estudis Espacials de Catalunya (IEEC), Carrer Gran Capit\`{a}, 08034 Barcelona, Spain\\
$^{3}$Harvard-Smithsonian Center for Astrophysics (CfA), 60 Garden Street, Cambridge, MA 02138, USA\\
$^{4}$Department of Physics \& Astronomy, Clemson University, Clemson, SC 29634, USA\\
$^{5}$Institute for Astronomy, University of Hawaii at Manoa, 2680 Woodlawn Drive, Honolulu, HI, 96822\\
$^{6}$Subaru Telescope, National Astronomical Observatory of Japan, 650 North A'ohoku Place, Hilo, HI, 96720, USA\\
$^{7}$Institut d'Astrophysique de Paris, UPMC et CNRS, UMR 7095, 98 bis bd Arago, F-75014 Paris, France
}

\date{Accepted XXX. Received YYY; in original form ZZZ}

\pubyear{2015}

\begin{document}
\label{firstpage}
\pagerange{\pageref{firstpage}--\pageref{lastpage}}
\maketitle

\begin{abstract}
We present a sample of 40 AGN in dwarf galaxies at redshifts $z \lesssim$ 2.4. The galaxies are drawn from the \textit{Chandra} COSMOS-Legacy survey as having stellar masses $10^{7}\leq M_{*}\leq3 \times 10^{9}$ M$_{\odot}$. Most of the dwarf galaxies are star-forming. After removing the contribution from star formation to the X-ray emission, the AGN luminosities of the 40 dwarf galaxies are in the range $L_\mathrm{0.5-10 keV} \sim10^{39} - 10^{44}$ erg s$^{-1}$. With 12 sources at  $z > 0.5$, our sample constitutes the highest-redshift discovery of AGN in dwarf galaxies. The record-holder is cid\_1192, at $z = 2.39$ and with $L_\mathrm{0.5-10 keV} \sim 10^{44}$ erg s$^{-1}$. One of the dwarf galaxies has $M_\mathrm{*} = 6.6 \times 10^{7}$ M$_{\odot}$ and is the least massive galaxy found so far to host an AGN. All the AGN are of type 2 and consistent with hosting intermediate-mass black holes (BHs) with masses $\sim 10^{4} - 10^{5}$ M$_{\odot}$ and typical Eddington ratios $> 1\%$. 
We also study the evolution, corrected for completeness, of AGN fraction with stellar mass, X-ray luminosity, and redshift in dwarf galaxies out to $z$ = 0.7. We find that the AGN fraction for $10^{9}< M_{*}\leq3 \times 10^{9}$ M$_{\odot}$ and $L_\mathrm{X} \sim 10^{41}-10^{42}$ erg s$^{-1}$ is $\sim$0.4\% for $z \leq$ 0.3 and that it decreases with X-ray luminosity and decreasing stellar mass. Unlike massive galaxies, the AGN fraction seems to decrease with redshift, suggesting that AGN in dwarf galaxies evolve differently than those in high-mass galaxies. Mindful of potential caveats, the results seem to favor a direct collapse formation mechanism for the seed BHs in the early Universe. 

\end{abstract}




\section{Introduction}
Supermassive black holes (SMBHs) reside at the center of most local massive galaxies (see reviews by \citealt{2013ARA&A..51..511K}; \citealt{2013ApJ...764..184M}) and power the most luminous quasars observed at $z\sim7$ (e.g., \citealt{2003AJ....125.1649F}; \citealt{2007AJ....134.2435W}; \citealt{2011Natur.474..616M}; \citealt{2013ApJ...779...24V}; \citealt{2015Natur.518..512W}); however, how these massive BHs form is not clearly understood. The existence of seed BHs of $100 < M_\mathrm{BH} < 10^{6}$ M$_{\odot}$ (or intermediate-mass BHs, IMBHs) at $z > 10$ has been invoked in order to explain the finding of SMBHs when the Universe was less than 1 Gyr old (e.g., see reviews by \citealt{2010A&ARv..18..279V,2012Sci...337..544V}; \citealt{2012NatCo...3E1304G}; \citealt{2016PASA...33...54R}; \citealt{2017IJMPD..2630021M}). Such seed BHs could form from the death of the first generation of (Population III) stars (e.g., \citealt{2004ARA&A..42...79B}), from direct collapse of inflowing dense gas in protogalaxies (e.g. \citealt{1994ApJ...432...52L}; \citealt{2006MNRAS.371.1813L}), or from mergers in dense stellar clusters (e.g., \citealt{1999A&A...348..117P}; \citealt{2009ApJ...694..302D}; \citealt{2016MNRAS.459.3432M}) and then grow via accretion and mergers to reach the $10^{9}$ M$_{\odot}$ of the high-z SMBHs in less than 1 Gyr (e.g., \citealt{2014GReGr..46.1702N}; \citealt{2016PASA...33....7J}). Alternatively, SMBHs at $z\sim5-6$ could directly form by mergers of protogalaxies (e.g., \citealt{2010Natur.466.1082M,2015ApJ...810...51M}; but see \citealt{2013MNRAS.434.2600F}). 

The presence of IMBHs at $z > 7$ is difficult to prove (e.g., \citealt{2015ApJ...808..139S}; \citealt{2016MNRAS.460.4003A}; \citealt{2016MNRAS.459.1432P}; \citealt{2016MNRAS.460.3143S}; \citealt{2017MNRAS.466.2131P}); however, observational evidence for their existence can be found in the local Universe as those 'leftover' seed BHs from the early Universe that did not grow into SMBHs (e.g., see review by \citealt{2017IJMPD..2630021M}). The first IMBHs were suggested in globular clusters, where several candidates have been found and stringent BH mass upper limits have been provided (e.g., \citealt{2005MNRAS.356L..17M}; \citealt{2012ApJ...750L..27S}; \citealt{2011A&A...533A..36L,2012A&A...542A.129L,2013A&A...552A..49L}; \citealt{2013A&A...554A..63F}; \citealt{2013ApJ...773L..31H}; \citealt{2014MNRAS.444...29L}; \citealt{2015AJ....150..120W,2016AJ....152...22W}; \citealt{2017Natur.542..203K}), and to explain those ultraluminous X-ray sources (ULXs) with X-ray luminosities above $5 \times 10^{40}$ erg s$^{-1}$, which are not easy to explain by stellar-mass BHs even when invoking super-Eddington accretion (e.g., \citealt{2009Natur.460...73F}; \citealt{2012MNRAS.423.1154S}; \citealt{2013MNRAS.436.3128M,2015MNRAS.448.1893M}; \citealt{2015MNRAS.454L..26H}). The three strongest IMBH candidates among ULXs have jet radio emission spatially coincident with the X-ray emission and are most likely the stripped nucleus of a dwarf galaxy undergoing a minor merger event with the ULX host galaxy (e.g., \citealt{2010ApJ...712L.107W,2017arXiv170404434W}; \citealt{2012ApJ...747L..13F}; \citealt{2012MNRAS.423.1309M}; \citealt{2013ApJ...768L..22S}; \citealt{2015MNRAS.448.1893M}; \citealt{2015ApJ...814....8K,2017ApJ...844L..21K}). The nuclei of dwarf galaxies are among the best places where to look for the leftover seed BHs, as dwarf galaxies have not significantly grown through merger and accretion processes and thus resemble the first galaxies that populated the early Universe. 

The search for low-mass BHs (M$_\mathrm{BH} \lesssim 10^{6}$ M$_{\odot}$) in dwarf galaxies is mostly based on the detection of X-ray emission (e.g., \citealt{2007ApJ...656...84G}; \citealt{2009ApJ...698.1515D}; \citealt{2011Natur.470...66R}; \citealt{2012ApJ...755..167D}; \citealt{2013ApJ...773..150S}; \citealt{2015ApJ...809L..14B,2017ApJ...836...20B}; \citealt{2015ApJ...805...12L}; \citealt{2015ApJ...798...38S}; \citealt{2016ApJ...831..203P}; \citealt{2017ApJ...837...48C}), in some cases spatially coincident with jet radio emission (e.g., \citealt{2014ApJ...787L..30R}; \citealt{2017ApJ...837...66N}), or the use of standard virial techniques to estimate the BH mass (e.g., \citealt{2004ApJ...607...90B}; \citealt{2004ApJ...610..722G,2007ApJ...670...92G}; \citealt{2005ApJ...632..799P}; \citealt{2013ApJ...775..116R}; \citealt{2015MNRAS.449.1526L}; \citealt{2016ApJ...831....2B}; \citealt{2017MNRAS.468L..97O}; see \citealt{2017IJMPD..2630021M} for a review). Additional searches in the infrared regime have yielded a few more candidates (e.g., \citealt{2007ApJ...663L...9S,2008ApJ...677..926S,2009ApJ...704..439S,2014ApJ...784..113S}; \citealt{2015MNRAS.454.3722S}; \citealt{2017A&A...602A..28M}). Most of these samples are however incomplete, very local ($z < 0.3$), skewed toward high Eddington ratios, or skewed toward type 1 AGN in the case of optical searches (e.g., \citealt{2004ApJ...610..722G,2007ApJ...670...92G}; \citealt{2013ApJ...775..116R}) which can hamper the detection of BHs lighter than $10^{5}$ M$_{\odot}$ if the size of the broad line region is controlled by BH mass (e.g., \citealt{2014MNRAS.437..740C}). \cite{2015ApJ...809L..14B} found an AGN with M$_\mathrm{BH} \sim 5 \times 10^{4}$ M$_{\odot}$ estimated using the virial technique, \cite{2014ApJ...782...55Y} performed a study of four low Eddington ratio sources, and \cite{2016ApJ...831..203P} searched for AGN in dwarf galaxies out to $z < 1$. Yet, these studies include very few sources. To circumvent the biases mentioned above, in \cite{2016ApJ...817...20M} we performed an X-ray stacking analysis of $\sim$ 50,000 dwarf galaxies selected in the COSMOS field making use of the recently completed \textit{Chandra} COSMOS-Legacy survey (\citealt{2016ApJ...819...62C}). We found that a population of IMBHs with X-ray luminosities $\sim 10^{39}-10^{40}$ erg s$^{-1}$ does exist in dwarf galaxies out to $z = 1.5$, and that their detection beyond the local Universe is most likely hampered by their low luminosity and mild obscuration unless deep surveys like the \textit{Chandra} COSMOS-Legacy are used. 

The \textit{Chandra} COSMOS-Legacy survey is a combination of the C-COSMOS survey (1.8 Ms, \citealt{2009ApJS..184..158E}) with 2.8 Ms of new \textit{Chandra} observations, thus covering a total of 2.2 deg$^{2}$ with a exposure time of 4.6 Ms. Nearly all the X-ray sources in the \textit{Chandra} COSMOS-Legacy survey have optical and near-infrared counterparts (\citealt{2016ApJ...817...34M}) from either \textit{Hubble}, \textit{Spitzer}, Subaru, Canada-France-Hawaii Telescope, Magellan, VLT, or VISTA observations and the field has additional been covered by \textit{Herschel}, \textit{GALEX}, \textit{XMM-Newton}, \textit{NuSTAR} and the VLA. The COSMOS survey constitutes thus the largest survey with a complete, deep ($i_\mathrm{AB} \sim 27$), multiwavelength dataset. 

In this paper we make use of the large area covered and the depth of COSMOS-Legacy to investigate the presence of accreting BHs in dwarf galaxies with stellar masses $10^{7}\leq M_{*}\leq3 \times 10^{9}$ M$_{\odot}$ (i.e., comparable or less massive than the Large Magellanic Cloud) beyond the local Universe. We find a total of 40 dwarf galaxies out to $z = 2.4$ hosting AGN of type 2, with a wide range of accreting rates (from sub-Eddington to super-Eddington accretion) and BH masses consistent with IMBHs. This constitutes the largest sample of IMBHs beyond the nearby Universe, circumventing the biases of previous samples, and allows us to study the fraction of AGN in dwarf galaxies and its evolution with redshift, X-ray luminosity and stellar mass down to X-ray luminosities of $\sim10^{41}$ erg s$^{-1}$ and $z$ = 0.7. The sample and X-ray analysis are presented in Section~\ref{sample}, while the results obtained are reported and discussed in Section~\ref{results}. Final conclusions are provided in Section~\ref{conclusions}. Throughout the paper we adopt a $\Lambda$CDM cosmology with parameters $H_{0}=70$ km s$^{-1}$ Mpc$^{-1}$, $\Omega_{\Lambda}=0.73$ and $\Omega_{m}=0.27$.

\section{Sample and Analysis}
\label{sample}
The sample of dwarf galaxies is drawn from a parent sample of $\sim$2300 X-ray-selected type 2 AGN out to $z \sim$3 (\citealt{2017ApJ...841..102S}) in the new \textit{Chandra} COSMOS-Legacy survey, which contains 4016 X-ray point sources down to a flux limit of 2.2 $\times 10^{-16}$, 1.5 $\times 10^{-15}$, and 8.9 $\times 10^{-16}$ erg cm$^{-2}$ s$^{-1}$ at 20\% completeness and in the 0.5-2 keV (soft), 2-10 keV (hard), and 0.5-10 keV (full) bands, respectively (\citealt{2016ApJ...819...62C}). The optical and infrared counterparts, spectroscopic ($z_\mathrm{spec}$) and photometric ($z_\mathrm{phot}$) redshifts, and X-ray properties are described in detail in \cite{2016ApJ...819...62C} and \cite{2016ApJ...817...34M}. Most of the sources with a $z_\mathrm{spec}$ have an spectroscopic accuracy $>99.5\%$, while those with a less reliable $z_\mathrm{spec}$ (spectroscopic accuracy $<99.5\%$) still have $\Delta z$ = $\frac{|z_\mathrm{spec}-z_\mathrm{phot}|}{1+z_\mathrm{spec}} < 0.1$ (see \citealt{2016ApJ...817...34M}). For these sources, we take the spectroscopic redshift as in \cite{2016ApJ...817...34M}. The $z_\mathrm{phot}$ have been obtained from the spectral energy distribution (SED) of the sources following the procedure described in \cite{2011ApJ...742...61S}: the COSMOS field has been observed in 31 different bands, so the analysis of the SED is equivalent to low-resolution spectroscopy with a an accuracy of $\sigma_{(\Delta z/(1+z_\mathrm{spec}))}$ = 0.03. This method produces a nominal value of the $z_\mathrm{phot}$, corresponding to the maximum of the redshift probability distribution function (PDF). The galaxy properties (mass, age, star formation rate ($SFR$), and galaxy type) are derived by performing a multi-wavelength analysis of the SEDs from far-infrared (500 $\mu$m), when available, to near-ultraviolet (2300 \AA) and using Bayesian statistics. The full details of the SED fitting can be found in \cite{2017ApJ...841..102S}. In short, the host galaxy properties are derived via 3-component SED fitting decomposition using a nuclear dust torus model (\citealt{2004MNRAS.355..973S}), a galaxy model (\citealt{2003MNRAS.344.1000B}) and starburst templates (\citealt{2001ApJ...556..562C}; \citealt{2002ApJ...576..159D}). The galaxy model templates are generated using a Chabrier initial mass function and exponentially decaying star formation histories with characteristic times ranging from $\tau = 0.1$ to 30 Gyr, and constant star formation. A PDF for galaxy parameters such as stellar mass and $SFR$ is then built to estimate the most representative value for each parameter. The $SFR$ is estimated by combining the contributions from UV and total IR luminosity ($L_\mathrm{8-1000 \mu m}$). In all cases the galaxy light dominates the optical emission while the AGN emission peaks at the mid-IR, when available. The mid-IR emission of these dwarf galaxies is however often contaminated by star formation; therefore, the SED of the dwarf galaxies is not dominated by the AGN.

Combining optical spectroscopic and photometric diagnostics, \cite{2016ApJ...817...34M} classify sources into Type 1 or unobscured AGN and type 2 and/or obscured AGN (see \citealt{2016ApJ...817...34M} for details on the classification). Of the type 2 and/or obscured AGN sample, \cite{2017ApJ...841..102S} measured a stellar mass for 2267 sources. From this last sample, we select dwarf galaxies as having the peak of the stellar mass PDF $10^{7}\leq M_{*}\leq3 \times 10^{9}$ M$_{\odot}$. We obtain as a result 51 dwarf galaxies with type 2 AGN, 45\% of which (23 sources) have spectroscopic redshifts. Twenty-four out of the 28 sources with photometric redshifts have narrow PDFs and therefore small uncertainties on $z$, with average 1$\sigma$ error $\langle \Delta z \rangle \sim$ 0.02. The remaining 4 sources have wide PDFs and/or two peaks in the PDF, and the $z_\mathrm{phot}$ are therefore poorly constrained, with 1$\sigma$ errors $\Delta z > 1$. For this reason, we decide not to take into account these 4 objects in our analysis. This reduces the sample to 47 dwarf galaxies. The average 1$\sigma$ errors on the stellar mass and $SFR$ derived from the PDF are $\langle \Delta$ log $M_{*} \rangle \sim$ 0.4 M$_{\odot}$ and $\langle \Delta SFR \rangle \sim$ 1 M$_\mathrm{\odot}$ yr$^{-1}$, respectively.

\subsection{ULXs in dwarf galaxies}
\label{ULX}
ULXs are typically associated with star-forming galaxies (as inferred from the X-ray luminosity function, e.g., \citealt{2011ApJ...741...49S}; \citealt{2012MNRAS.419.2095M}) and are observed to form preferentially in low-metallicity environments and low-metallicity galaxies (e.g., \citealt{2009MNRAS.400..677Z}; \citealt{2009MNRAS.395L..71M,2010MNRAS.408..234M}; \citealt{2013ApJ...769...92P}; \citealt{2014MNRAS.441.2346B}). The occurrence of ULXs in dwarf galaxies, which have typically high star formation rates and low metallicities, is thus expected to be higher than in more massive galaxies (e.g., \citealt{2008ApJ...684..282S}; \citealt{2011MNRAS.416.1844W}). 

\begin{figure*}
 \includegraphics[width=\textwidth]{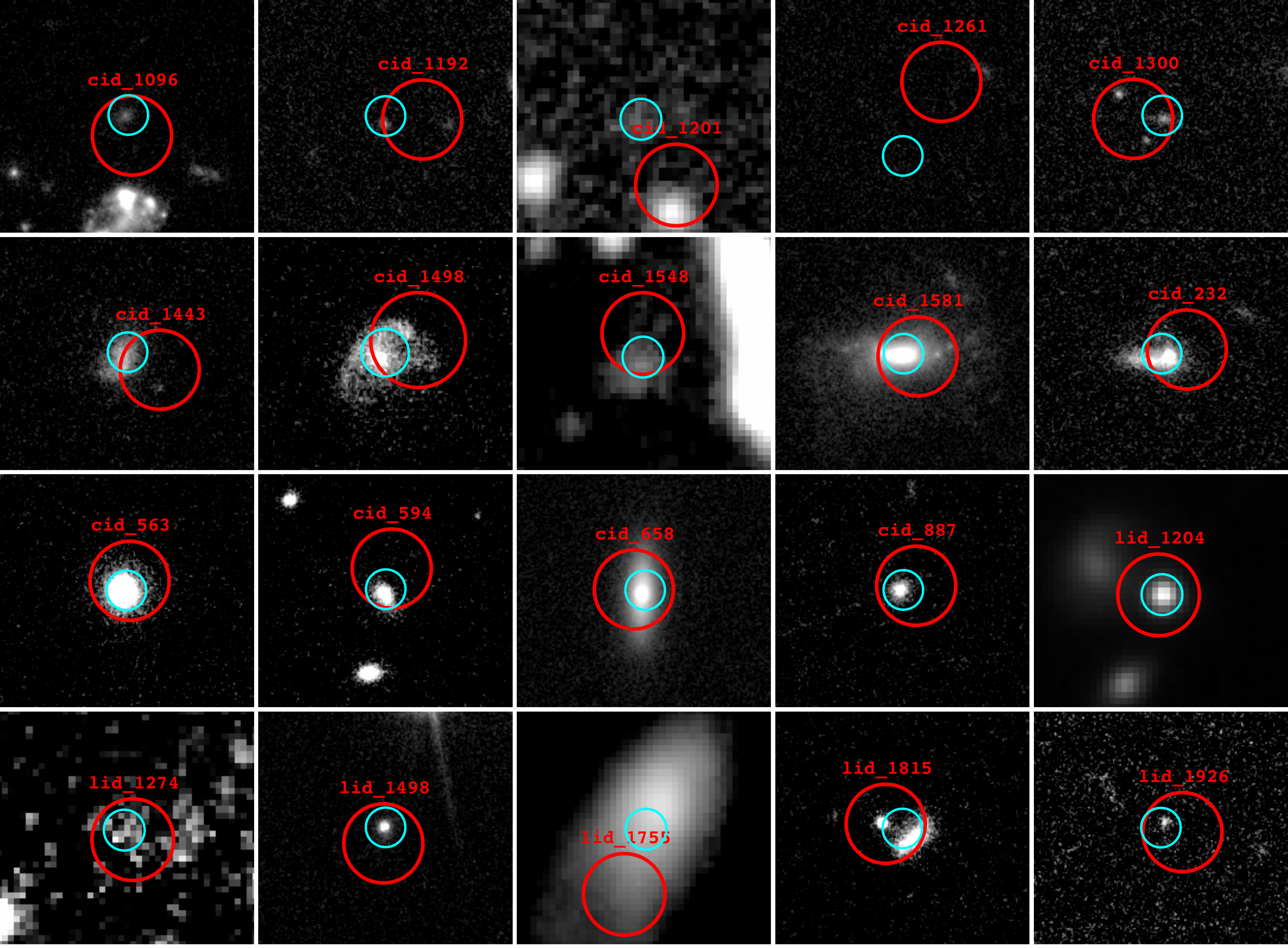}
\protect\caption[completeness]{\textit{Hubble Space Telescope} and Subaru i-band images of the dwarf galaxies studied in this work. The radius of the red circles denotes a \textit{Chandra} positional error of 1 arcsec. The cyan circles mark the optical position with a positional error of radius 0.5 arcsec.}
\label{mosaic}
\end{figure*}

\begin{figure*}
\includegraphics[width=\textwidth]{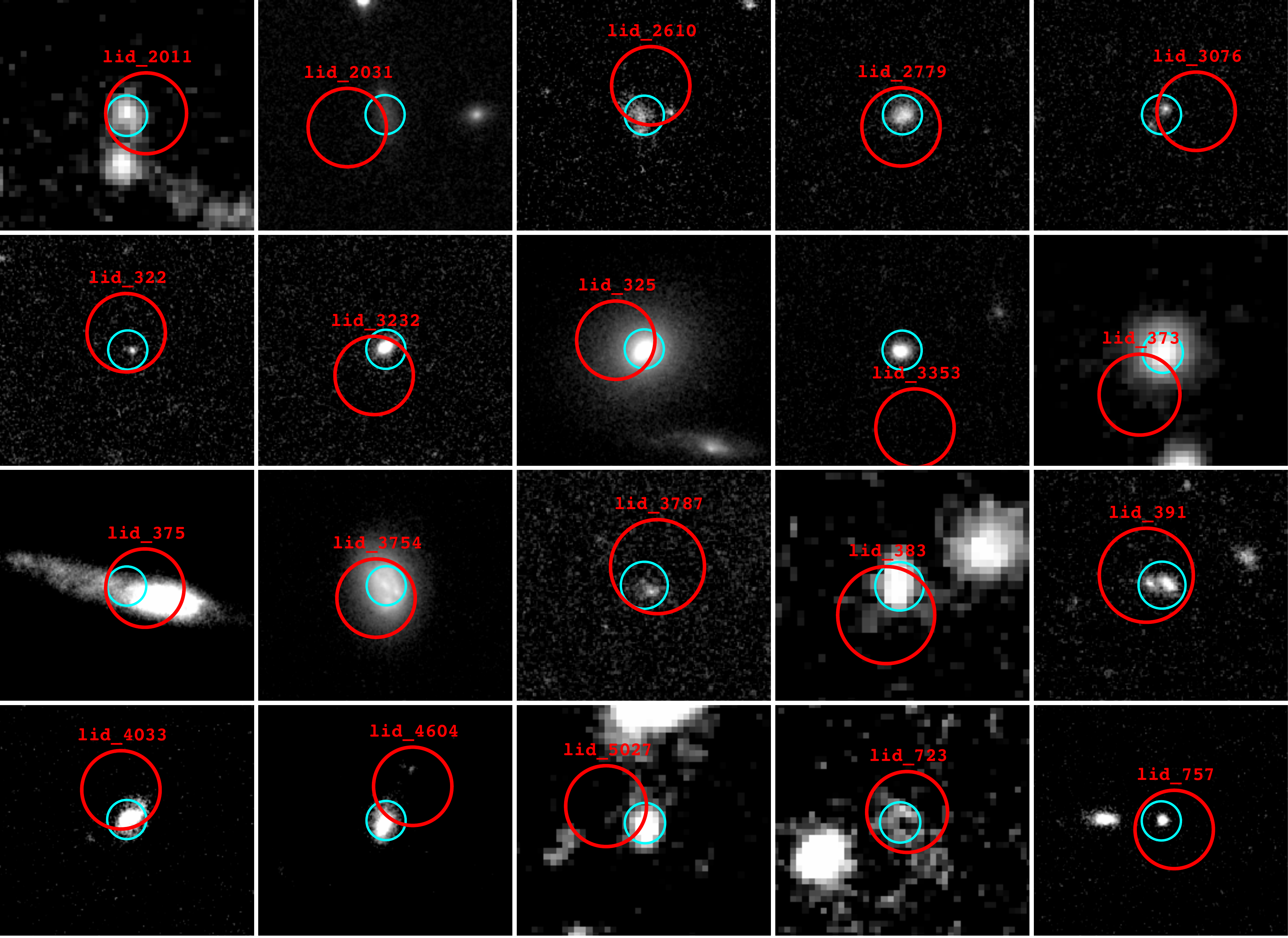}
\protect\caption[completeness]{Same caption as in Fig.~\ref{mosaic}.}
\label{mosaic2}
\end{figure*}

To investigate the presence of ULXs among the sample of 47 dwarf galaxies we apply the main criterium from \cite{2010A&A...514A..85M} for selecting ULXs in the COSMOS field: the distance between the X-ray and optical centroid has to be larger than 1.8 times the \textit{Chandra} positional error of the source. This yields 7 sources.
We note though that dwarf galaxies can have irregular morphologies and that it can be hard to define their dynamical center of mass, so that the 'central' BH may be anywhere within the core of the galaxy. To confirm the ULX nature of the 7 sources, we thus visually inspect the location of their \textit{Chandra} X-ray position on the optical images of the host galaxies. Six out of the 7 X-ray detections are clearly located in the outskirts (i.e. beyond the optical extension) of the host galaxies and not only off the optical center, which makes the criterium used above a reliable way of identifying ULXs. We thus remove the 7 sources from the sample in order to have a sample of AGN in dwarf galaxies as clean as possible. A detailed study of these 7 ULXs will be reported in a future work. There are four more sources in Figs.~\ref{mosaic}-\ref{mosaic2} that, visually, seem to be ULXs: cid\_1201, cid\_1261, lid\_1755, and lid\_3353. The optical/IR identification procedure is done in three bands (3.6 $\mu$m, K-band, and i-band) using a likelihood ratio (\citealt{2016ApJ...817...34M}). Very often this results in an association to a red source much brighter in the K-band that in the optical. This is the case of cid\_1201 and cid\_1261, whose counterpart is very faint in the i-band. For cid\_1261, the X-ray/optical position offset is due to the low source statistics in the X-ray band. The X-ray emission of lid\_1755 also seems slightly offset with respect to the optical center. However, the object does not accomplish the main ULX exclusion criterion from \cite{2010A&A...514A..85M}. In lid\_3353, the X-ray centroid is located between two potential X-ray sources. Its \textit{Chandra} exposure is relatively low and the point spread function (PSF) is quite poor (the offset from the centre of the observation is >5 arcmin). Yet, the separation of the optical counterpart from the northern X-ray blob is $<1$ arcsec. The X-ray/optical match for lid\_3353 is therefore also correct. Because of all the above, we do not identify cid\_1201, cid\_1261, lid\_1755, and lid\_3353 as ULXs. The final sample of dwarf galaxies with type 2 AGN and which constitutes the focus of this paper contains thus 40 sources.

\subsection{X-ray analysis}
\label{Xray}
Of the 40 AGN dwarf galaxies with X-ray emission in the \textit{Chandra} COSMOS-Legacy survey, 28 sources have more than 15 net counts in the 0.5-7 keV band. The X-ray fluxes and luminosities of these 28 galaxies have been obtained from a proper spectral fit with an absorbed power-law model. Sources with more than 30 net counts, which is the minimum number to perform a spectral analysis that is reliable (\citealt{2013MNRAS.431..978L}; \citealt{2016ApJ...830..100M}), have been analyzed in a separate work (\citealt{2016ApJ...830..100M}) while the spectra of sources having more than 15 and less than 30 net counts have been analyzed specifically for this work. All the fits are performed using the Cstat statistics, which is based on the Cash statistics (\citealt{1979ApJ...228..939C}) and is usually adopted for low-counts spectral fitting, since in principle it does not require counts binning to work. The fits are first performed fixing the power-law photon index to $\Gamma$=1.9, i.e., a typical AGN value, and leaving the intrinsic absorption N$_{\rm H}$ free to vary. We then repeat the fit leaving both $\Gamma$ and $N_{\rm H}$ free to vary, and we check if the fit is is significantly improved. We assume that a fit is significantly improved if Cstat$_\mathrm{old}$--Cstat$_\mathrm{new} > 2.71$ (see, e.g., \citealt{2006A&A...451..457T}; \citealt{2014MNRAS.443.1999B}), where Cstat$_\mathrm{old}$ and Cstat$_\mathrm{new}$ are the Cstat values for the fit without and with the photon index free to vary, respectively. In summary, 19 out of 28 sources have a best fit with fixed $\Gamma$=1.9, while the remaining 9 objects are best-fitted with a model having $\Gamma$ free to vary. For one source (cid\_563) the addition of an $\mathsf{APEC}$\footnote{\url{https://heasarc.gsfc.nasa.gov/xanadu/xspec/manual/node133.html}} model to describe thermal emission produces a significant improvement in the fit. We report the best-fit properties of the 28 dwarf galaxies with more than 15 net counts in the 0.5-7 keV band in Table~\ref{tab:xspec_prop}.

For the remaining 12 sources with less than 15 counts, the X-ray fluxes are taken from the recent \textit{Chandra} COSMOS-Legacy catalog (\citealt{2016ApJ...819...62C}) and converted to X-ray luminosities assuming $\Gamma$=1.4 (which takes into account the fact that obscuration produces a flattening in the observed spectrum) and Galactic $N_\mathrm{H}$= 2.6 $\times$ 10$^{20}$ cm$^{-2}$ (\citealt{2005A&A...440..775K}). The distribution of the full-band X-ray luminosity versus redshift for the 40 X-ray detected dwarf galaxies is plotted in Figure~\ref{Lxfull}. The solid line represents the 20\% completeness flux limit in the full band. Eight of the 40 AGN dwarf galaxies are not detected in the hard band above a maximum likelihood threshold $DET_{ML} = 10.8$ (corresponding to a Poisson probability of $P\simeq5 \times 10^{-5}$ that a detected source is actually a background fluctuation; \citealt{2016ApJ...819...62C}). We thus consider the 2-10 keV X-ray luminosities of these sources as upper limits in our analysis.

\begin{table*}
\begin{minipage}{\textwidth}
\centering
\caption{X-ray spectral fitting properties for the 28 dwarf galaxies with more than 15 net counts in the 0.5-7 keV band}
\label{tab:xspec_prop}
\begin{tabular}{ccccccccc}
\hline
\hline 
  ID & $z$ & cts & SNR & t$_\mathrm{exp}$ & Cstat/d.o.f & $N_{\rm H}$ & $\Gamma$ & Flux (2-10 keV)\\
  & & & & & (ks) & (10$^{22}$ cm$^{-2}$) & & (erg s$^{-1}$ cm$^{-2}$) \\
\hline
  cid\_232 & 0.165 & 81.98 & 9.26 & 163.20 &  18.15/25 & $<$0.27 & 1.36$_{-0.26}^{+0.39}$ & 8.09 $\times$ 10$^{-15}$\\
  cid\_563$^{\dagger}$ & 0.220 & 84.43 & 9.13 & 154.29 &   26.8/25 & $<$0.09 & 1.75$_{-0.38}^{+0.57}$ & 5.44 $\times$ 10$^{-15}$\\
  cid\_594 & 0.237 & 118.95 & 10.37 & 190.15 & 37.39/44 & 0.30$_{-0.24}^{+0.29}$ & 1.50$_{-0.28}^{+0.30}$& 1.24 $\times$ 10$^{-14}$\\
  cid\_658 & 0.121 & 252.45 & 13.76 & 175.94 & 79.31/75 & 0.92$_{-0.41}^{+0.51}$ & 1.85$_{-0.43}^{+0.48}$ & 2.18 $\times$ 10$^{-14}$ \\
  cid\_887 & 0.210 & 69.69 & 6.30 & 169.66 & 16.56/27 & 1.36$_{-0.71}^{+0.85}$ & 2.11$_{-0.60}^{+0.67}$ & 6.02 $\times$ 10$^{-15}$\\
  cid\_1096 & 0.263 & 20.24 & 2.36 & 196.62 & 30.85/20 & 0.85$_{-0.47}^{+0.61}$ & 1.9 & 2.30 $\times$ 10$^{-15}$\\
  cid\_1192 & 2.385 & 22.95 & 3.28 & 160.81 & 23.32/13 & $<$4.57 & 1.9 & 1.28 $\times$ 10$^{-15}$\\
  cid\_1261 & 1.750 & 25.24 & 4.21 & 127.33 & 27.0/24 & 10.04$_{-5.50}^{+8.48}$ & 1.9 & 3.90 $\times$ 10$^{-15}$\\
  cid\_1300 & 0.743 & 17.88 & 3.29 & 176.70 &  9.71/11 & 8.31$_{-3.98}^{+6.80}$& 1.9 & 2.73 $\times$ 10$^{-15}$\\
  cid\_1443 & 0.083 & 17.88 & 3.23 & 168.11 & 26.12/12 & $<$0.59 & 1.9 & 1.04 $\times$ 10$^{-15}$\\
  cid\_1498 & 0.354 & 24.95 & 4.20 & 159.85 & 14.74/16 &$<$1.19 & 1.9 & 1.21 $\times$ 10$^{-15}$\\
  cid\_1548 & 0.222 & 25.88 & 3.84 & 183.65 & 36.57/25 & 23.09$_{-6.90}^{+8.19}$ & 1.9 & 1.38 $\times$ 10$^{-14}$\\
  cid\_1581 & 0.196 & 43.85 & 7.20 & 172.06 &  5.14/16 & $<$0.11 & 1.9 & 2.78 $\times$ 10$^{-15}$\\
  lid\_322 & 0.215 & 73.39 & 10.23 & 154.11 &  37.78/26 & 2.85$_{-1.14}^{+1.28}$ & 2.65$_{-0.67}^{+0.74}$ & 8.43 $\times$ 10$^{-15}$\\
  lid\_373 & 1.130  & 20.49 & 6.12 & 47.79 & 14.1/8 & $<$4.81 & 1.9 & 3.85 $\times$ 10$^{-15}$\\ 
  lid\_375 & 0.08 & 21.61 & 3.54 & 184.79 & 11.86/23 & 0.83$_{-0.49}^{+0.60}$ & 1.9 & 2.35 $\times$ 10$^{-15}$\\
  lid\_383 & 0.568 & 21.35 & 5.12 & 78.51 & 16.53/15 & $<$0.80 & 1.9 & 2.44 $\times$ 10$^{-15}$\\
  lid\_391 & 0.505 & 148.92 & 13.18 & 174.37 & 45.54/43 & 1.51$_{-0.56}^{+0.63}$ & 1.73$_{-0.28}^{+0.30}$ & 1.51 $\times$ 10$^{-14}$\\
  lid\_723 & 1.280 & 52.63 & 8.89 & 169.56 & 20.05/17 & $<$3.87 & 0.96$_{-0.40}^{+0.47}$ & 9.26 $\times$ 10$^{-15}$\\
  lid\_757 & 0.367 & 135.77 & 12.46 & 125.22 & 46.13/40 & $<$0.09 & 1.43$_{-0.15}^{+0.16}$ & 1.65 $\times$ 10$^{-14}$\\
  lid\_1204 & 0.297 & 111.73 & 10.43 & 77.45 & 32.12/31 & $<$0.40 & 1.30$_{-0.21}^{+0.31}$ & 2.44 $\times$ 10$^{-14}$\\
  lid\_1274 & 0.976 & 16.23 & 3.48 & 159.76 & 16.57/10 & 11.80$_{-4.49}^{+6.97}$ & 1.9 & 3.65 $\times$ 10$^{-15}$\\
  lid\_1498 & 0.186 & 49.92 & 8.01 & 179.54 & 11.2/15 & $<$0.33 & 1.9 & 2.91 $\times$ 10$^{-15}$\\
  lid\_1755 & 0.054 & 42.51 & 7.13 & 44.57 & 7.16/12 & 0.93$_{-0.32}^{+0.38}$ & 1.9 & 1.52 $\times$ 10$^{-14}$\\
  lid\_1815 & 0.273 & 41.82 & 5.38 & 127.69 & 17.16/17 & 0.62$_{-0.35}^{+0.42}$ & 1.9 & 3.97 $\times$ 10$^{-15}$\\
  lid\_1926 & 1.282 & 21.05 & 3.87 & 161.80 &  9.03/11 & 3.93$_{-2.05}^{+2.55}$ & 1.9 & 2.29 $\times$ 10$^{-15}$\\
  lid\_3353 & 0.278 & 16.06 & 3.52 & 107.81 &  5.55/14 & 1.30$_{-1.05}^{+1.38}$ & 1.9 & 4.19 $\times$ 10$^{-15}$\\
  lid\_3787 & 0.085 & 32.62 & 5.17 & 178.30 &  7.55/10 & 5.15$_{-1.72}^{+2.65}$ & 1.9 & 4.65 $\times$ 10$^{-15}$\\
\hline
\hline
\end{tabular}
\end{minipage}
\raggedright
\smallskip\newline\small {\bf Column designation:} ID is the COSMOS-Legacy X-ray ID; cts are the 0.5-7 keV net counts; SNR is the signal-to-noise ratio in the 0.5-7 keV band; t$_\mathrm{exp}$  is the exposure time in the 0.5-7 keV band per source; Cstat/d.o.f. is the Cstat statistics best-fit value on the number of degrees of freedom; $N_{\rm H}$ is the intrinsic absorption; $\Gamma$ is the photon index; and the 2-10 keV flux is that computed from the X-ray spectrum. $^{\dagger}$ For cid\_563, the spectral fit includes an $\mathsf{APEC}$ model with $kT$ = 0.54$_{-0.4}^{+0.43}$. 
\end{table*}

\section{Results and Discussion}
\label{results}
\begin{figure}
 \includegraphics[width=0.5\textwidth]{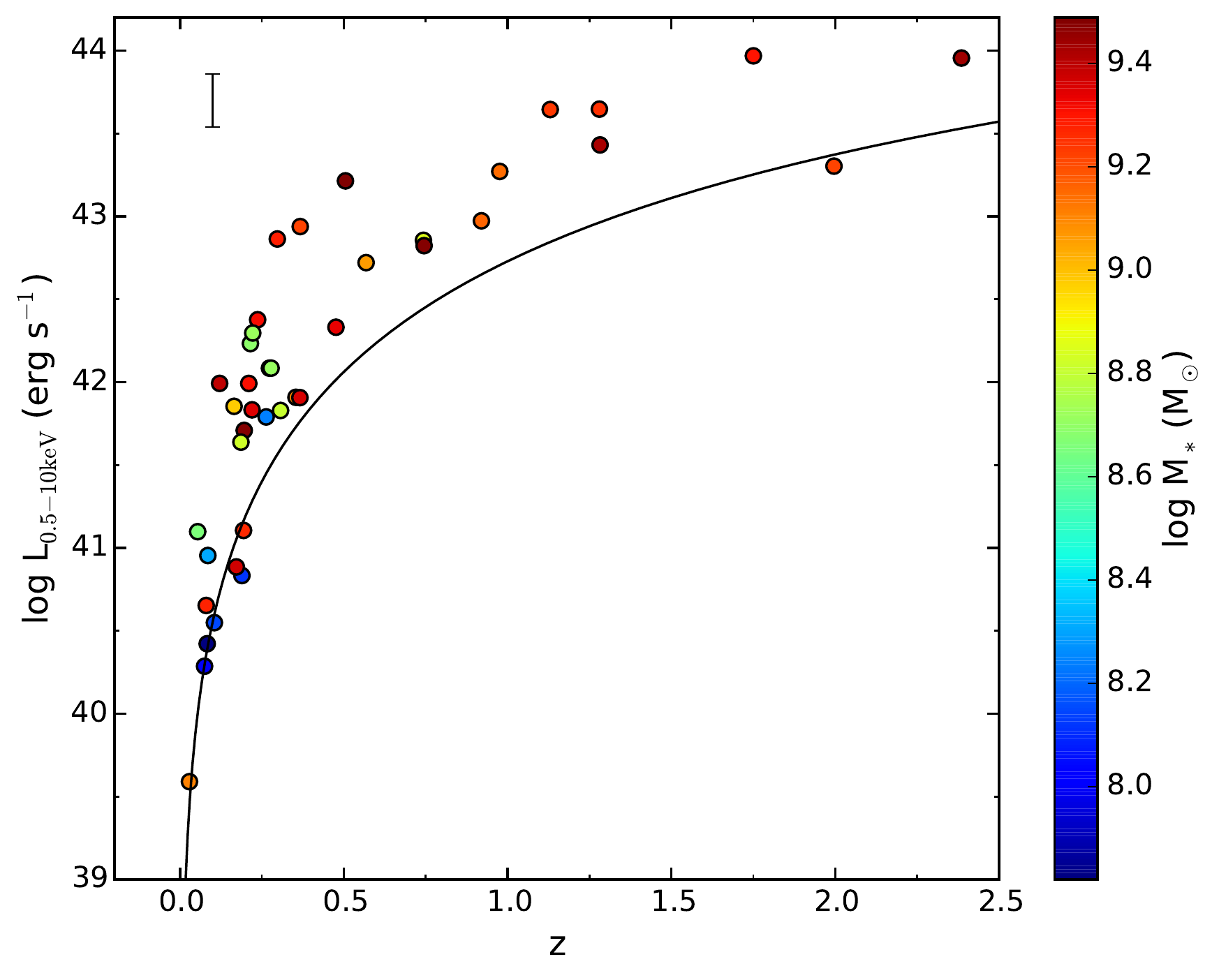}
\protect\caption[completeness]{Luminosity (0.5-10 keV) versus redshift for the sample of 40 dwarf galaxies hosting an AGN. The 20\% completeness flux limit in the full band is shown with a black solid line. The stellar mass is shown as a color bar.}
\label{Lxfull}
\end{figure}

\begin{figure}
\includegraphics[width=0.5\textwidth]{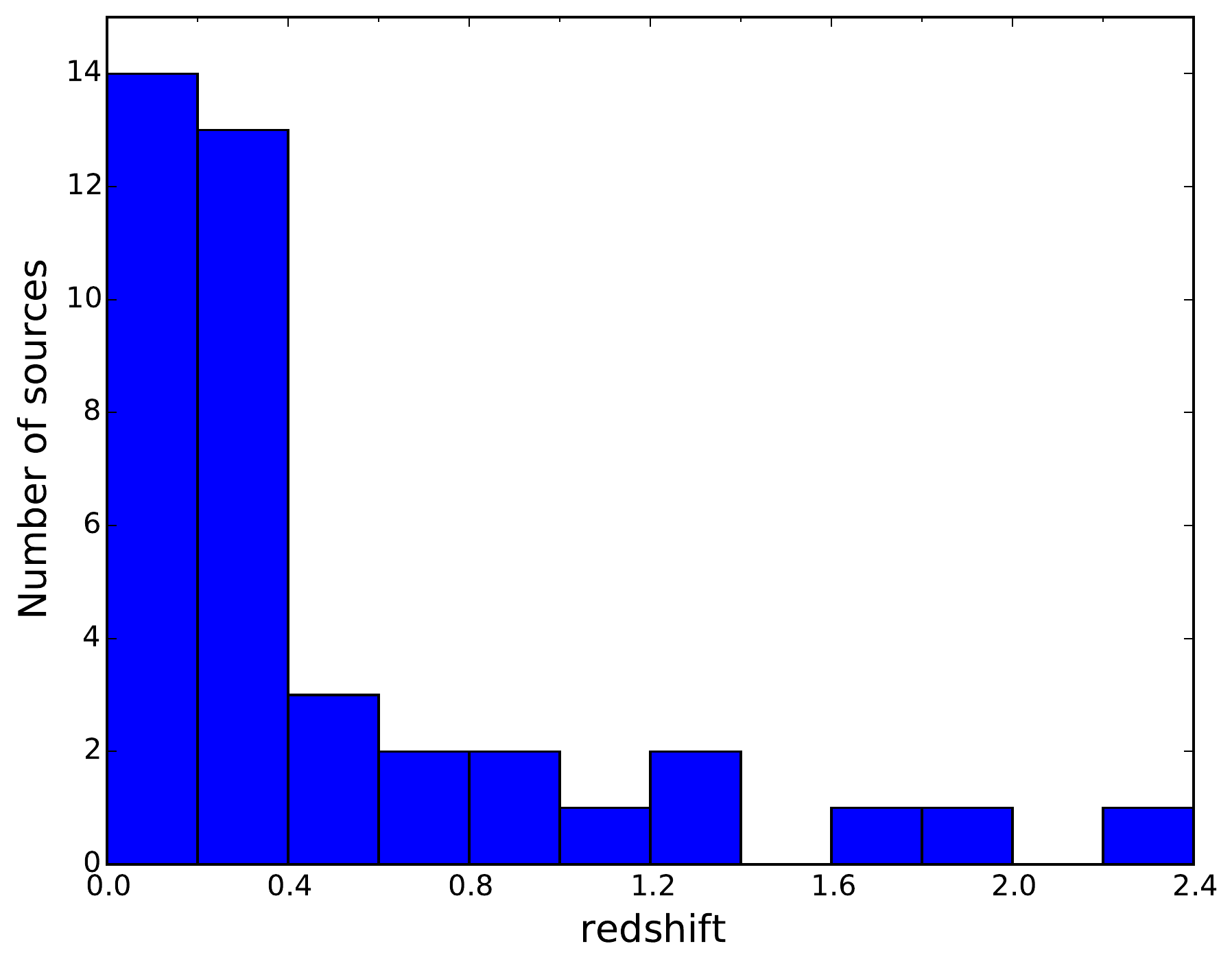}
\protect\caption[histoz]{Redshift distribution for the sample of 40 dwarf galaxies. There are 12 sources (30\%) with redshifts above 0.5.}
\label{histoz}
\end{figure}

The full-band K-corrected 0.5-10 keV X-ray luminosity of the 40 dwarf galaxies ranges from 3.5 $\times 10^{39}$ erg s$^{-1}$ to 9.3 $\times 10^{43}$ erg s$^{-1}$, with a mean value of L$_\mathrm{0.5-10 keV}$ = 1.0 $\times 10^{43}$ erg s$^{-1}$ (Figure~\ref{Lxfull}). Their redshift ranges from 0.03 to 2.39, with mean value of $z = 0.49$ (Figure~\ref{histoz}). This constitutes the highest-z sample of AGN in dwarf galaxies, which were so far limited to the local Universe (e.g., \citealt{2007ApJ...670...92G,2007ApJ...656...84G}; \citealt{2012ApJ...755..167D}; \citealt{2013ApJ...775..116R}; \citealt{2013ApJ...773..150S}) or out to $z < 1$ (\citealt{2016ApJ...831..203P}).  The range of stellar masses of the 40 dwarf galaxies goes from 6.6 $\times 10^{7}$ M$_{\odot}$ to 3.1 $\times 10^{9}$ M$_{\odot}$, and the $SFR$ from 2.3 $\times 10^{-3}$ M$_{\odot}$ yr$^{-1}$ to 46.7 M$_{\odot}$ yr$^{-1}$ (Table~\ref{table1}), from which we derive specific $SFR$s ($sSFR$) in the range 1.8 $\times 10^{-12}$ yr$^{-1}$ -- 2.6 $\times 10^{-8}$ yr$^{-1}$. The mean values of the sample are $M_\mathrm{*}$ = 1.4 $\times 10^{9}$ M$_{\odot}$, $SFR$ = 5.3 M$_{\odot}$ yr$^{-1}$, and $sSFR$ = 3.3 $\times 10^{-9}$ yr$^{-1}$. This classifies most of the galaxies as star-forming according to the tight correlation between stellar mass and $SFR$ recognized in the local Universe and intermediate redshifts (e.g., \citealt{2007ApJ...670..156D}; \citealt{2011ApJ...739L..40R}). The \textit{Hubble Space Telescope} and Subaru images of the 40 star-forming dwarf galaxies are shown in Figs.~\ref{mosaic}-\ref{mosaic2}.
The SED fitting provides a morphological classification (Ell-S0, Sa-Sc, Sd-Sdm, starburst) for each dwarf galaxy in our sample. We find that most galaxies (97.5\%) qualify as spiral and starburst, in agreement with being star-forming as inferred from their $sSFR$, while only one source is classified as early-type. The X-ray luminosity of the 40 dwarf galaxies is more than one order of magnitude higher than the typical X-ray luminosity of XRBs, suggesting that the emission is produced by BHs more massive than stellar. However, given the star-forming nature of most of the sources, the observed X-ray luminosity might be a mixed contribution of X-ray emission from XRBs, hot interstellar medium (ISM) gas, and nuclear (AGN) emission. The contribution from XRBs and hot ISM gas must thus be derived and removed from the detected X-ray emission in order to confirm the presence of AGN in the 40 dwarf galaxies. Sections~\ref{XRB}-\ref{hotISM} describe how this is performed. We will first use the hardness ratio as a proxy for investigating which type of X-ray emission we are observing. 

\subsection{Hardness ratio}
\label{HR}
\begin{figure}
\includegraphics[width=0.5\textwidth]{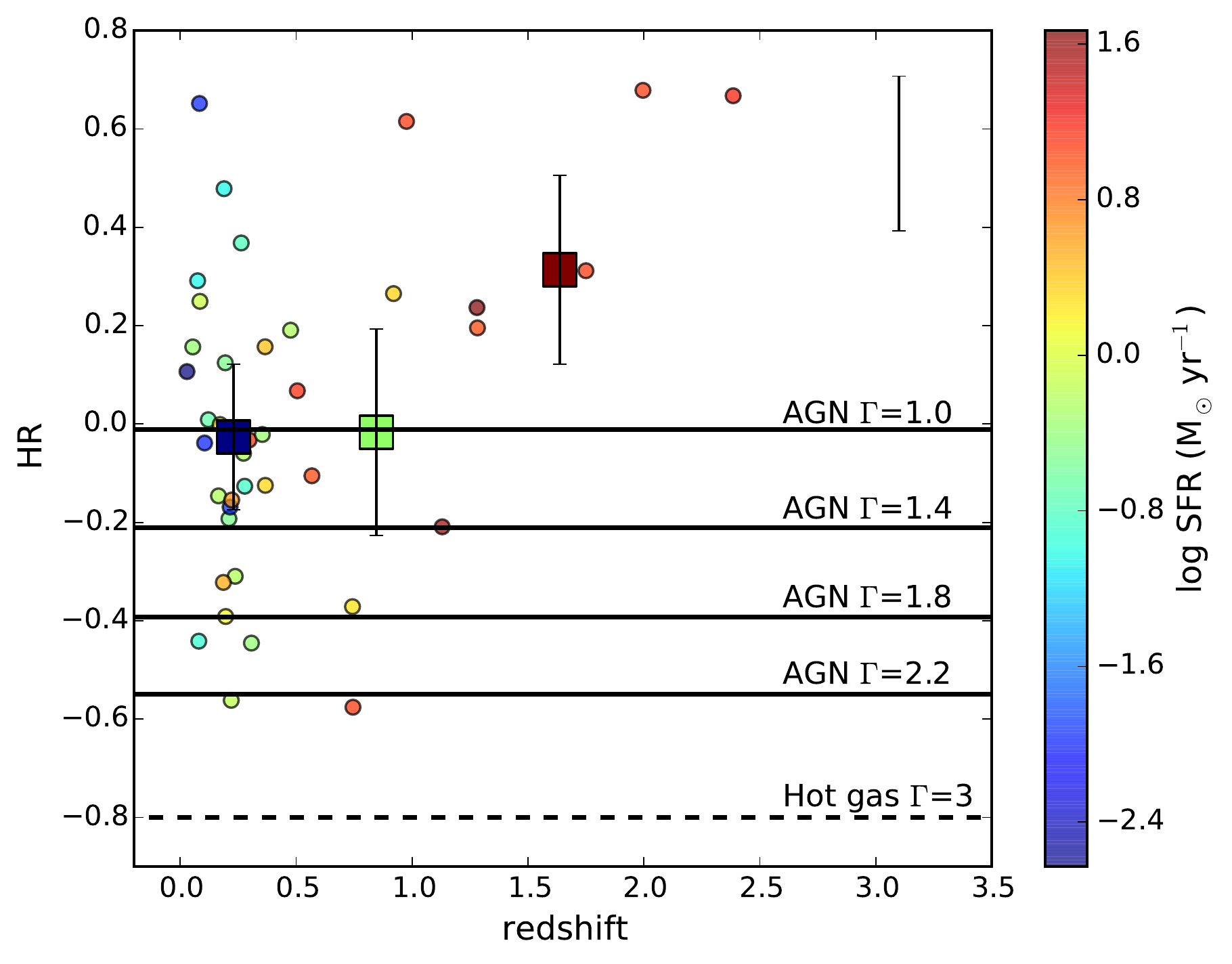}
\protect\caption[hardness]{Hardness ratio versus redshift and $SFR$ (color bar) for the 40 dwarf galaxies (filled dots). The dashed line indicates the $HR$ of thermal emission with $\Gamma = 3$, the solid lines represent the $HR$ of hard X-ray emission from moderately obscured AGN with photon index $\Gamma = 1$ and Galactic $N_\mathrm{H} = 2.6 \times 10^{20}$ cm$^{-2}$, and unobscured AGN with $\Gamma = 1.4$, $\Gamma = 1.8$ and $\Gamma = 2.2$ and Galactic $N_\mathrm{H}$. The error bar on the top right corner denotes the mean error of the 40 sources. The big squares are the mean $HR$ of the redshift bins: $0 < z < 0.6$, $0.6 \leq z < 1.0$, and $1.0 \leq z < 2.4$.}
\label{hardness}
\end{figure}

The X-ray hardness ratio ($HR$) is derived as $HR$ = ($H$-$S$)/($H$+$S$), where $H$ and $S$ are the count rates in the hard and soft bands, respectively. Obscured AGN ($N_\mathrm{H} > 10^{22}$ cm$^{-2}$) have typically $HR > 0$, while unobscured AGN showing unabsorbed soft spectra have $HR < -0.1$ (e.g., \citealt{2008A&A...490..905H}; \citealt{2012ApJS..201...30C}). Thermal emission from XRBs and hot ISM gas (associated with supernova remnants) have typically $HR \leq -0.8$. We show the distribution of $HR$ versus redshift for the sample of 40 dwarf galaxies in Figure~\ref{hardness}, where we plot also the expected $HR$ for thermal emission with $\Gamma = 3$ (dashed line) and for hard X-ray emission (solid lines) from either a population of XRBs or unobscured AGN (tracks at Galactic $N_\mathrm{H}$ with photon index $\Gamma = 1.4$, $\Gamma = 1.8$, and $\Gamma = 2.2$) and moderately obscured AGN (Galactic $N_\mathrm{H}$ and $\Gamma = 1$). We find that all the galaxies have a $HR$ above that of thermal emission from e.g., hot ISM gas but consistent with hard X-ray emission, which supports the AGN nature of the X-ray emission of all the galaxies in the sample. To divide the sources into obscured and unobscured AGN we use $HR = -0.1$. This results in 16 unobscured AGN (with $HR < -0.1$) and 24 obscured AGN. The fraction of obscured AGN is thus of 60\%, consistent with the $\sim50^{+17}_{-16}$ \% found by \cite{2016ApJ...819...62C} for the whole \textit{Chandra} COSMOS-Legacy catalog. The finding that some sources are unobscured despite of their optical classification as type 2 is not unexpected, as a source could be classified as type 2 in the optical not only when the source is obscured but also when the AGN is effectively less luminous than the galaxy light so that the optical SED is dominated by stellar emission (e.g., \citealt{2012ApJS..201...30C}; \citealt{2016ApJ...817...34M}).

In order to investigate any redshift evolution of the $HR$, we derive the mean values for three redshift bins: $0 < z < 0.6$, $0.6 \leq z < 1.0$, and $1.0 \leq z < 2.4$ (Figure~\ref{hardness}). A tendency towards higher $HR$ (i.e. higher column densities) for higher redshifts is observed, indicating a higher level of obscuration as $z$ increases. This is in agreement with an observed increase of the $SFR$ with $z$, which suggests a higher contribution of XRBs to the X-ray emission for higher redshifts, as was found by \cite{2016ApJ...817...20M} for a sample of stacked dwarf galaxies at $z < 1.5$. 

\subsection{Contribution from X-ray binaries} 
\label{XRB}
The contribution from XRBs will come predominantly from young, high-mass X-ray binaries (HMXBs), as low-mass X-ray binaries (LMXBs) typically populate early-type galaxies with very low $SFR$s and ages above a few Gyr (e.g., see review by \citealt{2006ARA&A..44..323F}). Since a small fraction of the dwarf galaxies (1 source) is of early-type, we consider both the contribution from HMXBs and LMXBs to the X-ray emission of the dwarf galaxies. Our sample spans a range of different redshifts and the XRB luminosity evolves with redshift due to a metallicity evolution, with lower metallicity at higher redshift, hence to estimate the contribution from HMXBs and LMXBs we use the evolution of their X-ray emission with metallicity ($Z$) and stellar age ($T$) derived by \cite{2013ApJ...776L..31F} from population synthesis simulations:

\begin{equation}
\label{FragosHMXBs}
\small
\mathrm{log}(L_\mathrm{X}^\mathrm{HMXBs}/SFR) = \beta_{0} + \beta_{1}Z + \beta_{2}Z^{2} + \beta_{3}Z^{3} + \beta_{4}Z^{4}  
\end{equation}
erg s$^{-1}$ M$_{\odot}$ yr, where $0 \leq Z \leq$ 0.025 and in the 2-10 keV band $\beta_{0}$ = 40.28 $\pm$ 0.02, $\beta_{1}$ = -62.12 $\pm$ 1.32, $\beta_{2}$ = 569.44 $\pm$ 13.71, $\beta_{3}$ = -1833.80 $\pm$ 52.14, and $\beta_{4}$ = 1968.33 $\pm$ 66.27.

\begin{equation}
\label{FragosLMXBs}
\begin{split}
\mathrm{log}(L_\mathrm{X}^\mathrm{LMXBs}/M_{*}) = \gamma_{0} + \gamma_{1}\mathrm{log}(T/Gyr) + \gamma_{2}\mathrm{log}(T/Gyr)^{2} + \\
+ \gamma_{3}\mathrm{log}(T/Gyr)^{3} + \gamma_{4}\mathrm{log}(T/Gyr)^{4} \ \mathrm{erg\ s}^{-1} 10^{10}\ \mathrm{M_{\odot}}
\end{split}
\end{equation}
where $0 \leq T \leq$ 13.7 Gyr and in the 2-10 keV band $\gamma_{0}$ = 40.276 $\pm$ 0.014, $\gamma_{1}$ = -1.503 $\pm$ 0.016, $\gamma_{2}$ = -0.423 $\pm$ 0.025, $\gamma_{3}$ = 0.425 $\pm$ 0.009, and $\gamma_{4}$ = 0.136 $\pm$ 0.009.

The 1$\sigma$ errors on the stellar mass and $SFR$ derived from the PDF are also included in the computation of the $L_\mathrm{2-10 keV}^\mathrm{HMXBs}$ and $L_\mathrm{2-10 keV}^\mathrm{LMXBs}$. To derive and subtract the XRB contribution ($L_\mathrm{2-10 keV}^\mathrm{XRB}$=$L_\mathrm{2-10 keV}^\mathrm{HMXBs}$+$L_\mathrm{2-10 keV}^\mathrm{LMXBs}$) to the detected 0.5-10 keV luminosity of the 40 dwarf galaxies, the $L_\mathrm{2-10 keV}^\mathrm{XRB}$ is converted to the 0.5-10 keV band assuming $\Gamma = 1.4$ (which is a good model for XRB emission; see e.g., \citealt{2006ApJ...645...95H}), $N_\mathrm{H} = 2.6 \times 10^{20}$ cm$^{-2}$ (see Section~\ref{Xray}), and applying the corresponding K-correction factor. We find that the observed $L_\mathrm{0.5-10 keV}$ is more than $\sim6\sigma$ larger than expected from the $L_\mathrm{XRB}$ derived from the correlations with $Z$ and $T$ for all the sources and that the contribution from XRBs to the 0.5-10 keV X-ray luminosity is below $\sim$16\%. In the hard band, we find that the $L_\mathrm{2-10 keV}$ is $>$ 11$\sigma$ above that expected from XRBs and that less than 9\% of the XRB emission contributes to the observed 2-10 keV X-ray emission. 

\begin{figure}
\includegraphics[width=0.5\textwidth]{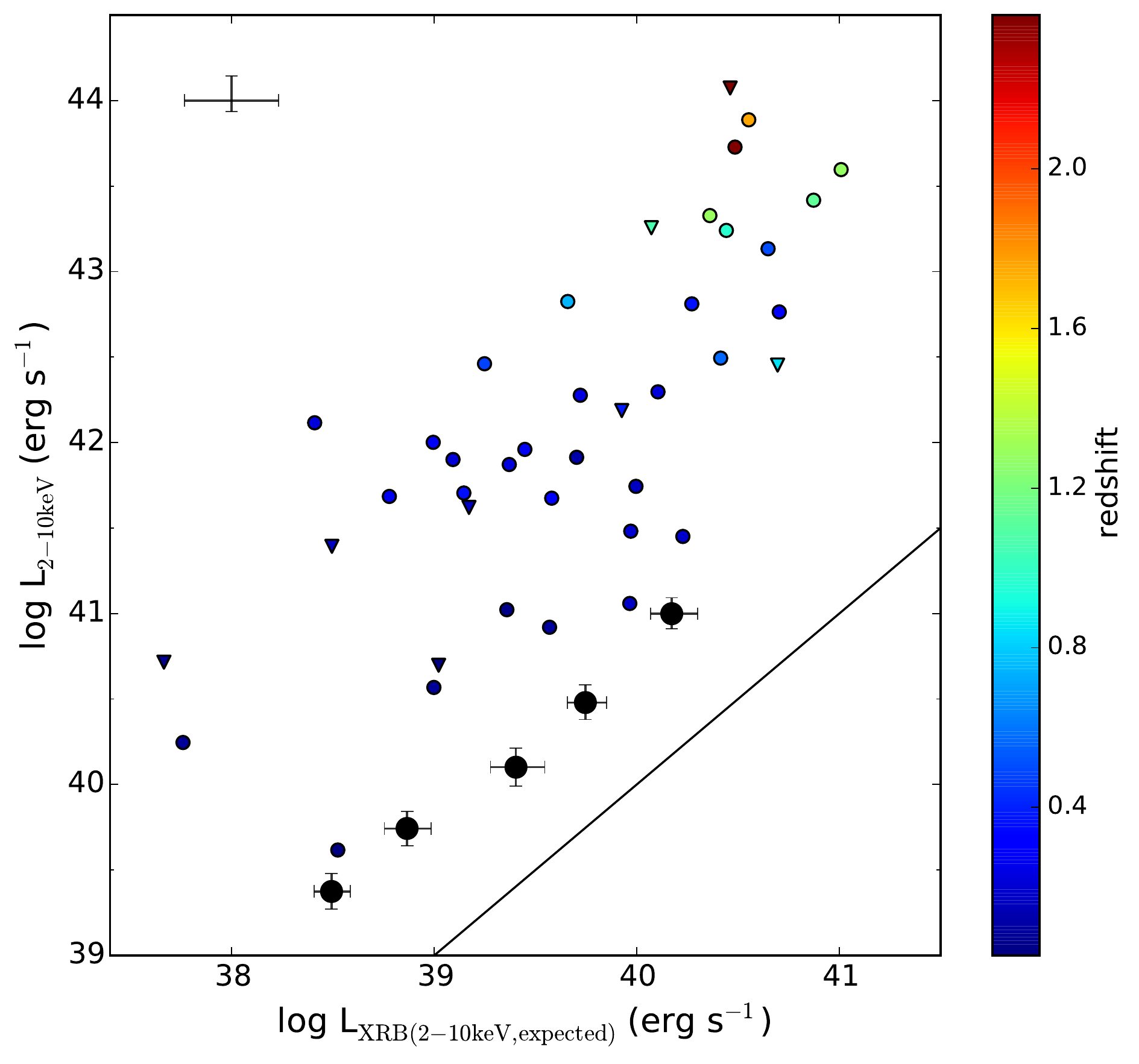}
\protect\caption[SFR]{2-10 keV observed X-ray luminosity of the 40 dwarf galaxies versus expected X-ray luminosity from LMXBs and HMXBs (derived using the evolution of the XRB emission with metallicity and stellar age from \citealt{2013ApJ...776L..31F}). The grey line shows a one-to-one correlation. The 1$\sigma$ error bar accounts for the uncertainties on the X-ray fluxes, the $SFR$s, the M$_{*}$, and the errors on the $\beta$ and $\gamma$ factors of the \cite{2013ApJ...776L..31F} correlations. Detections in the 2-10 keV are shown as circles and upper limits as inverted triangles. The $L_\mathrm{2-10 keV}$ is $> 11\sigma$ above the correlation for all the sources. The stacked X-ray luminosities of the sample of X-ray undetected dwarf galaxies from \cite{2016ApJ...817...20M} are overplotted as big black dots distributed in five redshift bins ($0 < z \leq 0.3$, $0.3 < z \leq 0.5$, $0.5 <  z \leq 0.7$, $0.7 < z \leq 1$, and $1 < z \leq 1.5$).}
\label{SFR}
\end{figure}

The observed $L_\mathrm{2-10 keV}$ versus the hard X-ray luminosity expected from XRBs according to the evolution of $L_\mathrm{X}^\mathrm{HMXBs}/SFR$ and $L_\mathrm{X}^\mathrm{LMXBs}/M_{*}$ with $Z$ and $T$ (eqs.~\ref{FragosHMXBs}-\ref{FragosLMXBs}) is plotted in Figure~\ref{SFR}. In the figure we also include the stacked 2-10 keV X-ray luminosities of the sample of non-X-ray detected dwarf galaxies from \cite{2016ApJ...817...20M}, which fill up the gap space of X-ray luminosities $L_\mathrm{2-10 keV} \sim 10^{39}-10^{40}$ erg s$^{-1}$ where no detections are found because of the flux limit of the \textit{Chandra} COSMOS-Legacy survey. 

Another correlation that takes into account the redshift evolution of the XRB emission is that found by \cite{2016ApJ...825....7L} for a population of normal galaxies at $z \sim$ 0-7:
\begin{equation}
\label{Lehmer2016}
\small
L_\mathrm{2-10 keV}^\mathrm{XRB} = \alpha_{0} (1+z)^{\gamma} M_{*} + \beta_{0} (1+z)^\delta SFR
\end{equation}
in erg s$^{-1}$, where log $\alpha_{0}$ = 29.30 $\pm$ 0.28, log $\beta_{0}$ = 39.40 $\pm$ 0.08, $\gamma$ = 2.19 $\pm$ 0.99, $\delta$ = 1.02 $\pm$ 0.22, the stellar mass term is proportional to LMXBs, the $SFR$ term is proportional to HMXBs, and the scatter is of 0.17 dex. We note that those galaxies with $L_\mathrm{2-10 keV} < 3 \times 10^{42}$ erg s$^{-1}$ are considered in \cite{2016ApJ...825....7L} as star-forming galaxies without an AGN, hence many of the dwarf galaxies in our sample would be included in the computation of the \cite{2016ApJ...825....7L} relation as lacking an AGN. Despite of this, using the \cite{2016ApJ...825....7L} relation we find that the observed $L_\mathrm{0.5-10 keV}$ is more than $\sim6\sigma$ larger than expected from the $L_\mathrm{XRB}$ and that the contribution from XRBs to the 0.5-10 keV X-ray luminosity is below $\sim$10\%. 

Using the scaling of $L_\mathrm{2-10 keV}^\mathrm{XRB}$ with M$_{*}$ and $SFR$ from \cite{2010ApJ...724..559L} for luminous star-forming galaxies and no redshift evolution:
\begin{equation}
\small
L_\mathrm{2-10 keV}^\mathrm{XRB} = (9.05 \pm 0.37) \times 10^{28} M_{*} + (1.62 \pm 0.22) \times 10^{39}\ SFR
\end{equation}
and which has a scatter of 0.34 dex, the observed $L_\mathrm{0.5-10 keV}$ would be more than $\sim22\sigma$ larger than expected from the $L_\mathrm{XRB}-M_\mathrm{*}-SFR$ correlation for all the sources and the contribution from XRBs to the 0.5-10 keV X-ray luminosity would be below $\sim$4\%. In the hard band, the $L_\mathrm{2-10 keV}$ would be more than $\sim32\sigma$ above that expected from star-forming galaxies and the contribution from XRBs to the 2-10 keV X-ray luminosity would be below $\sim$3\%. Using the $L_\mathrm{XRB}-M_\mathrm{*}-SFR$ correlation with no redshift evolution yields thus an X-ray excess significantly higher than that obtained from the \cite{2013ApJ...776L..31F} and \cite{2016ApJ...825....7L} relations (e.g., $\gtrsim22\sigma$ versus $\gtrsim6\sigma$ in the 0.5-10 keV band). The differences between these correlations and their implication for AGN searches are being explored by Fornasini et al. (in preparation).

There are several caveats that concern the XRB contribution estimate, all of which were investigated in detail in \cite{2016ApJ...817...20M}. First, we note that the relation between $SFR$ and X-ray luminosity becomes non-linear in the low $SFR$ regime (\citealt{2004MNRAS.351.1365G}), which could decrease the contribution from XRBs for those sources with the lowest $SFR$s. This issue was explored for the stacked sample of dwarf galaxies (\citealt{2016ApJ...817...20M}), finding that the differences obtained when using a non-linear relation of the form $L_\mathrm{XRB} \propto SFR^{1/(\alpha -1)}$ were consistent within the errors. Even if significant, the effect of such non-linearity would be to increase the X-ray excess, hence giving further support to the presence of AGN in the sample of dwarf galaxies studied here. 

Second, the most significant caveat that can affect the XRB contribution estimate is that the metallicity for high redshift and highly star-forming galaxies is expected to be lower than the solar value assumed in the SED fitting (e.g., \citealt{2010A&A...521L..53L}; \citealt{2013ApJ...763....9Y}; \citealt{2014ApJ...792...75Z}), which can result in a higher contribution of HMXBs to the integrated X-ray luminosity (e.g., \citealt{2013ApJ...776L..31F}; \citealt{2016ApJ...825....7L}; \citealt{2016ApJ...817...20M}). To evaluate this effect, we estimate the XRB contribution using the \cite{2013ApJ...776L..31F} correlations assuming that the metallicity is half solar ($Z = 0.0067$). We find that in this case the $L_\mathrm{2-10 keV}$ would still be more than $\sim7\sigma$ above that expected from XRBs and that the contribution from XRBs to the 2-10 keV X-ray luminosity would be below $\sim$14\%. 


\begin{table*}
\begin{minipage}{\textwidth}
\centering
\caption{Host galaxy properties. The average 1$\sigma$ errors on $M_\mathrm{*}$ and $SFR$ are $\sim$0.4 M$_{\odot}$ and  $\sim$1 M$_\mathrm{\odot}$ yr$^{-1}$, respectively.}
\label{table1}
\begin{tabular}{lccccccccc}
\hline
\hline 
  ID 		&      RA(J2000)    &     DEC(J2000) &  z            &  log $M_\mathrm{*}$	&  i 		&  log $SFR$ 				   &  log $L_\mathrm{hot}$ & log $L_\mathrm{XRBs}$	&	log $L_\mathrm{AGN}$ (0.5-10 keV)	\\
  	 	 & 		(deg)  	&  (deg) 		& 		&  (M$_\mathrm{\odot}$)	&  (mag)	& (M$_\mathrm{\odot}$ yr$^{-1}$) &              (erg s$^{-1}$)		&     	 (erg s$^{-1}$)	&        	(erg s$^{-1}$)		\\
    (1)	    	&   (2)     	  &     (3)   	&   	(4)     &  	(5)				&	  (6)   & 		(7)   	               		     &  		  (8)  				 & 		(9)  		&			(10)  \\    
 \hline
 cid\_1096  &  150.115620  &  1.951608  &  0.26$^{+0.01}_{-0.01}$$^{p}$ &  8.24  &  23.41  &     -0.78  &  38.24$\pm$0.01  &  38.78$\pm$0.02  &   41.8$\pm$0.0  \\
 cid\_1192  &  149.885210  &  2.372808  &  2.38$^{+0.37}_{-0.22}$$^{p}$ &  9.44  &  24.81  &      1.20  &  40.6$\pm$0.2  &  40.5$\pm$0.2  &   43.9$\pm$0.0  \\
 cid\_1201  &  149.869540  &  2.357775  &   2.00$^{+0.95}_{-0.85}$$^{p}$ &   9.22  &  25.93  &      1.06  &  40.5$\pm$0.2  &  40.5$\pm$0.2  &   43.3$\pm$0.2  \\
 cid\_1261  &  150.466510  &  2.298900  &  1.75$^{+0.53}_{-0.39}$$^{p}$ &     9.30  &  25.43  &      1.08  &  40.4$\pm$0.2  &   40.6$\pm$0.2  &   43.9$\pm$0.0  \\
 cid\_1300  &  150.101950  &  2.163916  &  0.74$^{+0.11}_{-0.03}$$^{p}$ &    8.85  &  23.90  &      0.25  &  39.41$\pm$0.08  &   39.66$\pm$0.08  &   42.9$\pm$0.0  \\
 cid\_1443  &  150.009250  &  2.390140  &         0.083$\pm$0.000$^{s}$ &    7.82  &  22.25  &     -1.96  &   37.00$\pm$0.04  &  37.76$\pm$0.05  &   40.4$\pm$0.0  \\
 cid\_1498  &  149.773580  &  2.141633  &         0.354$\pm$0.000$^{s}$ &    9.10  &  22.28  &     -0.36  &  38.69$\pm$0.04  &   39.15$\pm$0.04  &   41.9$\pm$0.0  \\
 cid\_1548  &  150.005845  &  2.295230  &         0.222$\pm$0.000$^{s}$ &    8.72  &  24.17  &      0.61  &  39.62$\pm$0.01  &  40.10$\pm$0.02  &   42.3$\pm$0.0  \\
 cid\_1581  &  150.628380  &  2.414759  &         0.196$\pm$0.000$^{s}$ &    9.48  &  20.15  &      0.13  &  39.12$\pm$0.05  &  39.97$\pm$0.02  &    41.7$\pm$0.0  \\
  cid\_232  &  149.988640  &  1.908585   &         0.165$\pm$0.000$^{s}$ &    8.97  &  22.29  &     -0.23  &  38.76$\pm$0.07  &  40.0$\pm$0.1  &   41.9$\pm$0.0  \\
  cid\_563  &  149.920540  &  2.543668   &          0.220$\pm$0.000$^{s}$ &     9.35  &  21.15  &     -0.17  &   40.98$^{\dagger}$  &   39.37$\pm$0.01  &   41.8$\pm$0.0  \\
  cid\_594  &  150.501346  &  2.354369   &         0.237$\pm$0.000$^{s}$ &     9.32  &  23.16  &     -0.23  &  38.78$\pm$0.07  &  39.72$\pm$0.07  &   42.4$\pm$0.0  \\
  cid\_658  &  150.336048  &  2.433799   &         0.121$\pm$0.000$^{s}$ &     9.39  &  20.11  &     -0.68  &  38.29$\pm$0.01  &  39.70$\pm$0.01  &   41.9$\pm$0.0  \\
  cid\_887  &  150.070390  &  2.286302   &          0.210$\pm$0.000$^{s}$ &     9.31  &  23.21  &     -0.52  &  38.48$\pm$0.07  &  39.09$\pm$0.06  &   41.9$\pm$0.0  \\
 lid\_1204  &  150.130571  &  1.505687   &   0.30$^{+0.01}_{-0.01}$$^{p}$ &     9.28  &  21.49  &      1.11  &  40.14$\pm$0.01  &  40.70$\pm$0.02  &   42.9$\pm$0.0  \\
 lid\_1274  &  150.058114  &  1.653284   &  0.98$^{+0.05}_{-0.18}$$^{p}$ &   9.14  &  26.33  &      1.06  &  40.28$\pm$0.03  &   40.44$\pm$0.03  &   43.3$\pm$0.0  \\
 lid\_1498  &  150.521084  &  1.880207   &  0.19$^{+0.01}_{-0.03}$$^{p}$ &    8.82  &  22.89  &      0.51  &  39.51$\pm$0.05  &  40.2$\pm$0.1  &   41.61$\pm$0.01  \\
 lid\_1755  &  150.865828  &  2.051638   &  0.05$^{+0.02}_{-0.01}$$^{p}$ &    8.66  &  19.92  &     -0.36  &  38.58$\pm$0.05  &  39.36$\pm$0.03  &   41.1$\pm$0.0  \\
 lid\_1815  &  149.739201  &  2.667379   &         0.273$\pm$0.000$^{s}$ &     8.83  &  22.59  &     -0.25  &  38.78$\pm$0.08  &  39.45$\pm$0.06  &   42.1$\pm$0.0  \\
 lid\_1926  &  150.328320  &  2.494110   &  1.28$^{+0.16}_{-0.03}$$^{p}$ &   9.43  &  25.35  &      0.99  &  40.3$\pm$0.1  &  40.4$\pm$0.1  &   43.4$\pm$0.0  \\
 lid\_2011  &  150.743505  &  2.456254   &         0.366$\pm$0.000$^{s}$ &     9.36  &  24.94  &      0.42  &  39.47$\pm$0.05  &  39.93$\pm$0.05  &   41.9$\pm$0.2  \\
 lid\_2031  &  150.511100  &  2.599287  &         0.105$\pm$0.000$^{s}$ &     8.15  &  22.34  &     -1.99  &  36.97$\pm$0.05  &   37.67$\pm$0.05  &  40.6$\pm$0.2  \\
 lid\_2610  &  149.781186  &  1.660789   &  0.74$^{+0.03}_{-0.02}$$^{p}$ &   9.48  &  23.73  &      1.07  &  40.23$\pm$0.02  &  40.70$\pm$0.07  &   42.8$\pm$0.1  \\
 lid\_2779  &  150.477240  &  1.872818   &  0.19$^{+0.02}_{-0.02}$$^{p}$ &    8.12  &  23.32  &     -1.05  &  37.94$\pm$0.01  &  38.50$\pm$0.02  &   40.8$\pm$0.2  \\
 lid\_3076  &  150.716114  &  2.479516   &         0.307$\pm$0.000$^{s}$ &     8.81  &  24.04  &     -0.38  &  38.65$\pm$0.09  &  39.6$\pm$0.2  &  41.8$\pm$0.1  \\
  lid\_322  &  150.105140  &  2.795295   &  0.21$^{+0.01}_{-0.02}$$^{p}$ &   8.69  &  24.75  &     -1.89  &  37.11$\pm$0.01  &  38.4$\pm$0.2  &   42.2$\pm$0.0  \\
 lid\_3232  &  150.717953  &  2.788256   &  0.48$^{+0.06}_{-0.04}$$^{p}$ &    9.34  &  23.31  &     -0.26  &  38.83$\pm$0.04  &  39.25$\pm$0.04  &  42.3$\pm$0.2  \\
  lid\_325  &  150.196073  &  2.821245   &         0.029$\pm$0.000$^{s}$ &   9.11  &  19.55  &     -2.63  &   36.3$\pm$0.02  &  38.52$\pm$0.04  &  39.5$\pm$0.2  \\
 lid\_3353  &  149.479629  &  2.206855   &         0.278$\pm$0.000$^{s}$ &     8.71  &  22.77  &     -0.86  &  38.16$\pm$0.08  &  39.00$\pm$0.07  &   42.1$\pm$0.0  \\
  lid\_373  &  149.921131  &  2.886119  &  1.13$^{+0.03}_{-0.02}$$^{p}$ &   9.24  &  23.23  &      1.51  &  40.76$\pm$0.05  &   40.87$\pm$0.05  &   43.6$\pm$0.0  \\
  lid\_375  &  149.957674  &  2.796841   &          0.080$\pm$0.000$^{s}$ &     9.27  &  19.60  &     -0.92  &  38.04$\pm$0.02  &  39.00$\pm$0.01  &   40.6$\pm$0.0  \\
 lid\_3754  &  150.467448  &  1.865929   &         0.172$\pm$0.000$^{s}$ &     9.37  &  20.13  &      0.22  &   39.2$\pm$0.05  &   39.97$\pm$0.03  &  40.8$\pm$0.3  \\
 lid\_3787  &  150.612276  &  2.114749   &  0.09$^{+0.01}_{-0.01}$$^{p}$ &  8.31  &  22.82  &     -0.10  &  38.86$\pm$0.06  &  39.57$\pm$0.06  &   40.9$\pm$0.0  \\
  lid\_383  &  150.035418  &  2.963838   &    0.6$^{+0.2}_{-0.2}$$^{p}$ &    9.06  &  23.77  &      0.98  &   40.1$\pm$0.02  &  40.41$\pm$0.02  &   42.7$\pm$0.0  \\
  lid\_391  &  150.442403  &  2.763864   &         0.505$\pm$0.000$^{s}$ &   9.49  &  23.50  &      1.14  &  40.24$\pm$0.04  &  40.65$\pm$0.04  &   43.2$\pm$0.0  \\
 lid\_4033  &  150.408872  &  2.509424   &         0.194$\pm$0.000$^{s}$ &     9.26  &  22.98  &     -0.52  &  38.47$\pm$0.07  &  39.17$\pm$0.07  &   41.1$\pm$0.2  \\
 lid\_4604  &  150.039578  &  1.779599   &  0.07$^{+0.01}_{-0.01}$$^{p}$ &    8.02  &  21.61  &     -1.04  &  37.91$\pm$0.06  &  39.02$\pm$0.09  &  40.3$\pm$0.2  \\
 lid\_5027  &  150.777616  &  2.420784   &  0.92$^{+0.03}_{-0.08}$$^{p}$ &    9.16  &  24.54  &      0.34  &  39.54$\pm$0.07  &  40.1$\pm$0.2  &  42.9$\pm$0.1  \\
  lid\_723  &  149.553471  &  2.376666   &  1.28$^{+0.34}_{-0.06}$$^{p}$ &    9.25  &  25.34  &      1.67  &  40.95$\pm$0.01  &  41.01$\pm$0.01  &   43.6$\pm$0.0  \\
  lid\_757  &  149.533635  &  2.244504   &  0.37$^{+0.01}_{-0.02}$$^{p}$ &    9.22  &  23.78  &      0.29  &  39.35$\pm$0.09  &   40.3$\pm$0.1  &   42.9$\pm$0.0  \\
 \hline
\hline
\end{tabular}
\end{minipage}
\raggedright
\smallskip\newline\small {\bf Column designation:}~(1) \textit{Chandra} COSMOS-Legacy X-ray ID, (2) right ascension, (3) declination, (4) redshift, $s$ = spectroscopic, $p$ = photometric, (5) stellar mass, (6) i-band magnitude, (7) star formation rate, (8) 0.5-2 keV X-ray luminosity expected from hot ISM gas estimated using the correlation from \cite{2012MNRAS.426.1870M}, (9) 2-10 keV X-ray luminosity expected from XRBs estimated using the correlation from \cite{2013ApJ...776L..31F}, (10) 0.5-10 keV AGN X-ray luminosity after removing the contribution from XRBs and hot ISM gas. $^{\dagger}$ For cid\_563, the $L_\mathrm{0.5-2 keV}^\mathrm{hot}$ comes from the fit of the X-ray spectrum with a thermal component in addition to the power-law model.
\end{table*}

\subsection{Contribution from hot ISM gas}
\label{hotISM}
In star-forming galaxies, the X-ray emission might include some contribution from hot ISM gas in addition to XRBs (e.g., \citealt{2013MNRAS.428.2085L}). To estimate the hot ISM contribution to the $L_\mathrm{0.5-10 keV}$ of the 40 dwarf galaxies we use the correlation between $SFR$ and diffuse gas X-ray luminosity ($L_\mathrm{0.5-2 keV}^\mathrm{hot}$) from \cite{2012MNRAS.426.1870M}:
\begin{equation}
\label{eqhotISM}
L_\mathrm{0.5-2 keV}^\mathrm{hot} = (8.3 \pm 0.1) \times 10^{38}\ SFR\ \mathrm{(M}_{\odot} \mathrm{yr}^{-1}\mathrm{)} 
\end{equation}
which has a scatter of 0.34 dex. We convert the $L_\mathrm{0.5-2keV}^\mathrm{hot}$ to the 0.5-10 keV band assuming a power-law index of $\Gamma=3$ (which is a good representation of a thermal model with temperature $\sim$0.7-1 keV) and apply the corresponding K-correction factor. Using equation~\ref{eqhotISM} we find that the $L_\mathrm{0.5-10 keV}$ is more than $\sim34\sigma$ above that expected from hot ISM gas for all the sources and that the contribution of hot ISM to the 0.5-10 keV luminosity is below $\sim$3\%. This is reinforced by the finding that the spectral fitting of those sources with more than 15 counts does not improve by the addition of a thermal component to the power-law model (see Sect.~\ref{Xray}). Only for one source, cid\_563, are we able to estimate the $L_\mathrm{0.5-2 keV}^\mathrm{hot}$ directly from the fit of the X-ray spectrum (see Table~\ref{tab:xspec_prop}), finding log $L_\mathrm{0.5-2 keV}^\mathrm{hot}$ = 40.98 erg s$^{-1}$ with a large uncertainty (of nearly one order of magnitude). Even when considering these uncertainties, the $L_\mathrm{0.5-2 keV}^\mathrm{hot}$ of cid\_563 derived from the spectral fitting is significantly higher than its log $L_\mathrm{0.5-2 keV}^\mathrm{hot}$ = 38.75 erg s$^{-1}$ estimated from the \cite{2012MNRAS.426.1870M} correlation. Taking the $L_\mathrm{0.5-2 keV}^\mathrm{hot}$ derived from the spectral fitting, the $L_\mathrm{0.5-10 keV}$ of cid\_563 is $\sim5\sigma$ above that from the hot ISM gas and the contribution of the hot ISM to the 0.5-10 keV luminosity for cid\_563 is $\sim$22\%. This is two orders of magnitude higher than the contribution from the hot ISM gas that is obtained for cid\_563 using equation~\ref{eqhotISM}. When feasible, spectral fitting rather than the $L_\mathrm{X,hot}- SFR$ correlation should thus be preferably performed to estimate the contribution from hot gas to the X-ray emission. We use the log $L_\mathrm{0.5-2 keV}^\mathrm{hot}$ = 40.98 erg s$^{-1}$ obtained from the spectral fitting to remove the contribution from hot gas to the X-ray emission for cid\_563 in the next section.

\subsection{AGN emission}
\label{nuclear}
We derive the AGN emission by subtracting the contribution of XRBs and diffuse hot gas emission from the detected 0.5-10 keV band luminosity. Given that these contributions were not very significant ($\lesssim$16\%), we find that the AGN luminosities in the 0.5-10 keV band still range from $\sim$ 3.5 $\times 10^{39}$ erg s$^{-1}$ to 9.3 $\times 10^{43}$ erg s$^{-1}$ in the redshift range $z = $0.03 to 2.39. The finding of AGN in dwarf galaxies at such high redshifts constitutes an unprecedented discovery: \cite{2016ApJ...817...20M} found that a population of IMBHs exists in low-mass galaxies out to $z < 1.5$, while the so-far redshift record-holder for an AGN in a low-mass galaxy was source ID 31097, with $z = 0.53$ and $L_\mathrm{0.5-7 keV}$ = 1.3 $\times 10^{42}$ erg s$^{-1}$ (\citealt{2016ApJ...831..203P}). Thanks to the wide area and sensitivity of \textit{Chandra} COSMOS-Legacy, we find 12 AGN dwarf galaxies with $z > 0.5$ (Figure~\ref{histoz}), the new record-holder being cid\_1192 ($z = 2.39$, $L_\mathrm{0.5-10 keV}$ = 9.0 $\times 10^{43}$ erg s$^{-1}$). 

\begin{figure}
\includegraphics[width=0.5\textwidth]{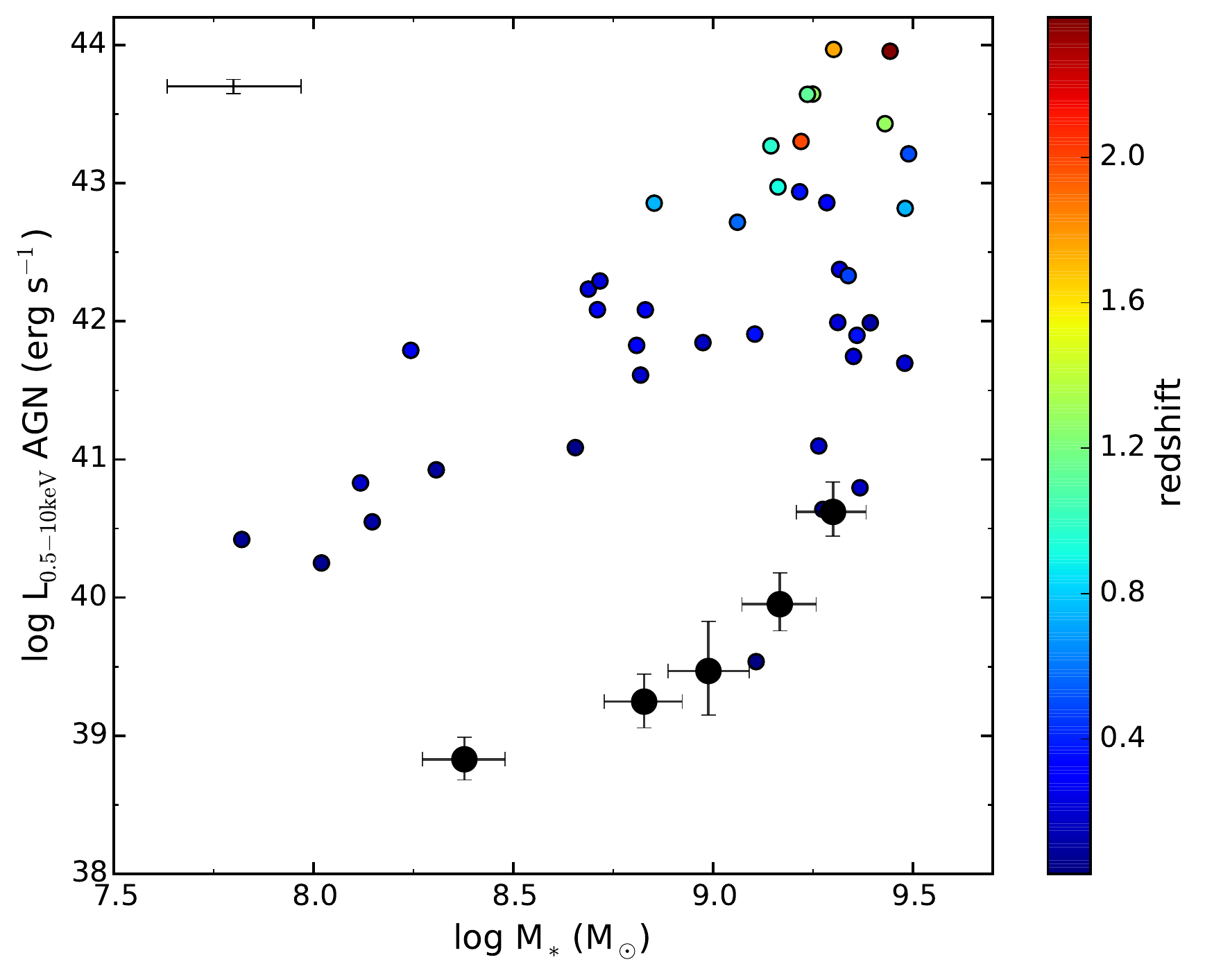}
\protect\caption[AGN]{Nuclear X-ray luminosity in the 0.5-10 keV band versus stellar mass for the 40 dwarf galaxies. The error bar on the top left corner denotes the mean error on $L_\mathrm{AGN}$ and $M_\mathrm{*}$ of the sources. The stacked X-ray luminosities of the sample of X-ray undetected dwarf galaxies from \cite{2016ApJ...817...20M} are overplotted as big black dots distributed in five redshift bins ($0 < z \leq 0.3$, $0.3 < z \leq 0.5$, $0.5 <  z \leq 0.7$, $0.7 < z \leq 1$, and $1 < z \leq 1.5$).}
\label{AGN}
\end{figure}

The variation of $L_\mathrm{AGN}$ (0.5-10 keV) with $M_\mathrm{*}$ and $z$ is shown in Figure~\ref{AGN}, where we also plot the stacked nuclear X-ray luminosity for five redshift bins out to $z < 1.5$ from \cite{2016ApJ...817...20M}. The new 40 detected AGN fill the region with $L_\mathrm{X} > 10^{40}$ erg s$^{-1}$ for log $M_\mathrm{*} <  9.5$, where similar bright ($L_\mathrm{X} > 10^{40}$ erg s$^{-1}$) AGN dwarf galaxies (e.g., from \citealt{2011Natur.470...66R}; \citealt{2013ApJ...773..150S}; \citealt{2014ApJ...787L..30R}; \citealt{2015ApJ...809L..14B,2017ApJ...836...20B}; \citealt{2015ApJ...798...38S}; \citealt{2016ApJ...831..203P}) would be also located. As reported by \cite{2016ApJ...817...20M}, the lack of high stellar mass sources with low X-ray luminosities observed in Figure~\ref{AGN} is due to the mass limit of the COSMOS optical/infrared survey, while the observed increase of $L_\mathrm{AGN}$ with stellar mass and redshift is due to the X-ray survey limit. The increase of $L_\mathrm{AGN}$ with $M_\mathrm{*}$ and $z$ could be also caused by the higher $SFR$ found at higher $z$, which hinders the measure of the AGN contribution, while at lower $z$ (and lower $SFR$) the AGN contribution can be more easily measured and is thus more significant (\citealt{2016ApJ...817...20M})

\subsection{Black hole mass and accretion rate}
\label{BHmass}
\begin{figure}
\includegraphics[width=0.5\textwidth]{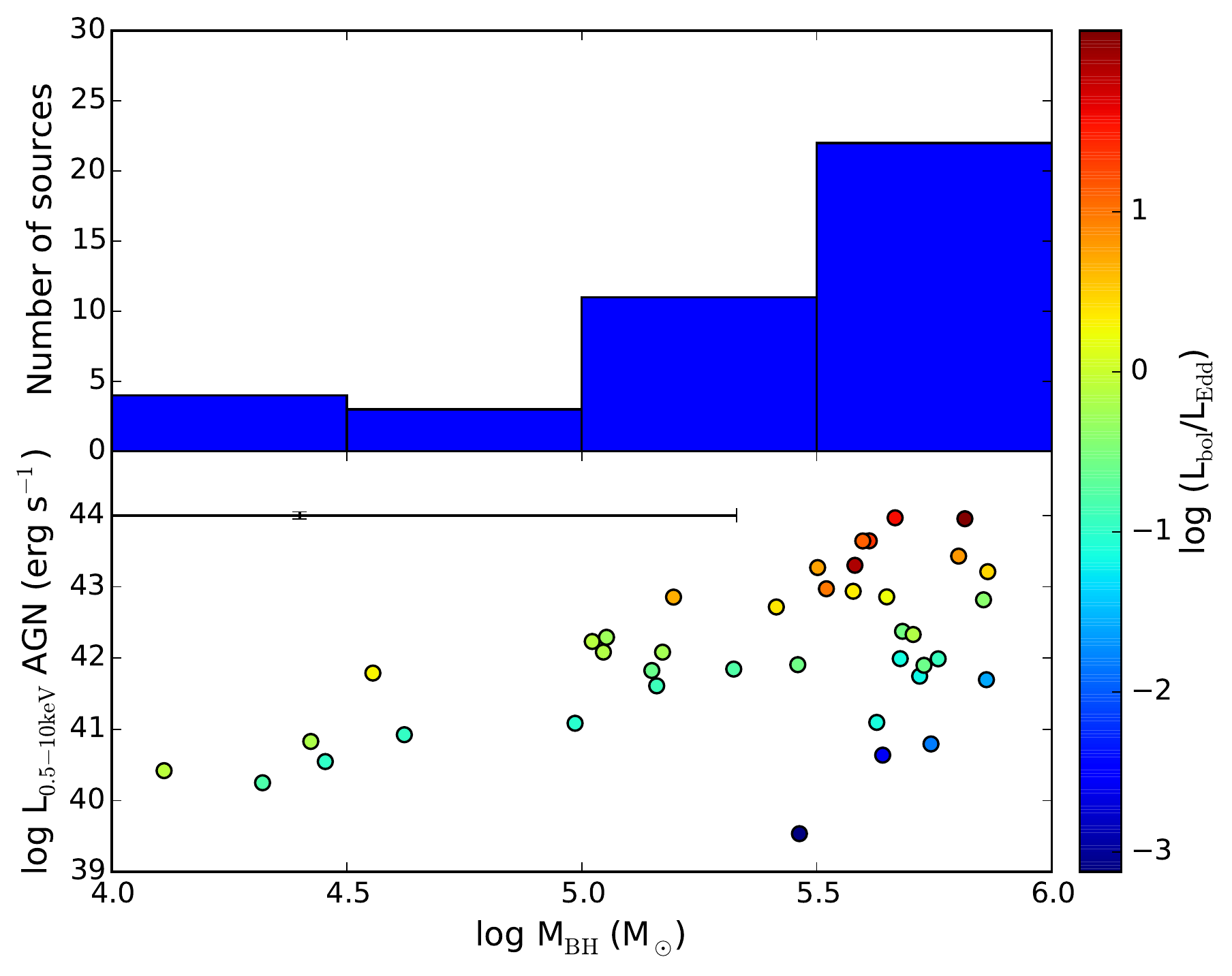}
\protect\caption[Mbh]{\textbf{Top:} Distribution of BH masses for the 40 dwarf galaxies. The $M_\mathrm{BH}$ have been estimated using the correlation from \cite{2015ApJ...813...82R}. \textbf{Bottom:} AGN luminosity in the 0.5-10 keV band versus BH mass. The error bar on the top left shows the mean error of the $L_\mathrm{AGN}$ and BH mass. All the sources have $M_\mathrm{BH} < 10^{6}$ M$_{\odot}$ and are thus consistent with being IMBHs. Given the error on the BH mass, those sources with log $M_\mathrm{BH} \geq 5$ M$_{\odot}$ cannot be discarded as possible SMBHs. The color bar denotes the Eddington ratio.}
\label{Mbh}
\end{figure}

To get an estimate of the BH mass of the 40 dwarf galaxies hosting AGN, we consider the correlation between BH mass and stellar mass found for local AGN (including dwarf galaxies) by \cite{2015ApJ...813...82R}:
log ($M_\mathrm{BH}/$M$_{\odot}$) = 7.45 $\pm$ 0.08 + (1.05 $\pm$ 0.11) log ($M_\mathrm{*}/10^{11}$ M$_{\odot}$), with a scatter of 0.55 dex. We find BH masses in the range $M_\mathrm{BH} = 1.3 \times 10^{4} - 7.3 \times 10^{5}$ M$_{\odot}$ with a mean value of $M_\mathrm{BH} = 3.3 \times 10^{5}$ M$_{\odot}$, indicating that all the sources are consistent with being IMBHs (see Figure~\ref{Mbh}). We note that, although the correlation includes a sample of dwarf galaxies, scaling relations are in general not well calibrated in the low-mass regime (e.g., \citealt{2018ApJ...855L..20M}) and thus their use for estimating BH masses in dwarf galaxies is not highly reliable. To quantify such unreliability and be as conservative as possible, we compute the uncertainty on the BH mass considering both the minimum and maximum $M_\mathrm{BH}$ provided by the 0.55 dex scatter and the minimum and maximum stellar mass derived from the PDF. As a result we find that, when considering these maximized uncertainties on the BH mass, 30 out of 40 of the AGN dwarf galaxies could be as well SMBHs.

To estimate the accretion rate we derive the Eddington ratio $l_\mathrm{Edd} = L_\mathrm{bol}/L_\mathrm{Edd}$, where $L_\mathrm{bol}$ is the bolometric luminosity derived from the SED fitting (\citealt{2017ApJ...841..102S}) and $L_\mathrm{Edd} = 1.3 \times 10^{38} \times M_\mathrm{BH}/M_{\odot}$ erg s$^{-1}$. The uncertainty on the Eddington ratio estimates is maximized by propagating the uncertainties obtained for the BH mass. The distribution of Eddington ratios is shown as a color bar in Figure~\ref{Mbh}. We find that the Eddington ratio ranges from $7.5 \times 10^{-4}$ to $1.4 \times 10^{2}$, with a median value of $l_\mathrm{Edd} = 0.6$. Albeit the large uncertainties, we find that most of the sources (95\%) have near- to super-Eddington accretion rates ($l_\mathrm{Edd} > 10^{-2}$), as commonly found in optically-selected samples of low-mass BHs in the local Universe for which the BH mass has been derived in more robust ways (e.g. \citealt{2004ApJ...610..722G,2007ApJ...670...92G}; \citealt{2012ApJ...755..167D}; \citealt{2013ApJ...775..116R}; \citealt{2014ApJ...782...55Y}; \citealt{2015ApJ...809L..14B,2017ApJ...836...20B}). Only two of the AGN dwarf galaxies are found to be accreting at sub-Eddington rates. The high Eddington rates could be explained if BH accretion occurs through short accretion episodes that can reach or even exceed the Eddington limit, as expected from models of early BH growth (e.g., \citealt{2005ApJ...633..624V}; \citealt{2014ApJ...784L..38M}; \citealt{2015MNRAS.451.1964S}; \citealt{2016MNRAS.458.3047P}; \citealt{2017MNRAS.472L.109A}).

Of particular interest is the finding of a subset of four dwarf galaxies hosting AGN with $M_\mathrm{BH} = (1.3 - 2.9) \times 10^{4}$ M$_{\odot}$ (Figure~\ref{Mbh}). They are at $z < 0.2$, their stellar masses range from $M_\mathrm{*} = 6.6 \times 10^{7}$ M$_{\odot}$ to 1.4 $\times 10^{8}$ M$_{\odot}$ and they are all near- to super-Eddington accreting ($l_\mathrm{Edd} = 0.8-1.0$). These sources have the lowest BH mass of the sample and, if the scaling relations hold in this mass regime (though see e.g., \citealt{2018ApJ...855L..20M}), they would be the lightest IMBHs ever found in dwarf galaxies (the lightest AGN so far reported is that of the dwarf galaxy RGG 118, with $M_\mathrm{BH} \sim 5 \times 10^{4}$ M$_{\odot}$; \citealt{2015ApJ...809L..14B}). One of them, with $M_\mathrm{*} = 6.6 \times 10^{7}$ M$_{\odot}$, represents as well the least massive galaxy so far known to host an AGN. A detailed study of these extraordinary IMBHs will be explored in a future work.

\subsection{Radio emission}
To investigate the presence of radio emission in the sample of dwarf galaxies, we make use of the VLA-COSMOS 3 GHz Large Project source catalog (\citealt{2017arXiv170309713S}), which consists of 384 hours of deep Karl G. Jansky VLA observations covering the entire 2 deg$^{2}$ COSMOS field at 10 cm (3 GHz) down to an rms of 2.3 $\mu$Jy beam$^{-1}$ and with a resolution of 0.75 arcsec. We find that 3 out of the 40 dwarf galaxies have a radio counterpart above the VLA-COSMOS threshold of 5.5$\sigma$ within a radius $<$ 1 arcsec from the \textit{Chandra} X-ray counterpart. The redshift of these three sources ranges from $z = 0.03$ to $z = 0.2$ and their Eddington ratios from $l_\mathrm{Edd} = 7 \times 10^{-4}$ to $l_\mathrm{Edd}$ = 0.8 (see Table~\ref{tab:radio}). 
We convert the 3 GHz radio fluxes to 1.4 GHz assuming a radio spectral index $\alpha = 0.8$ (typical of star-forming galaxies, \citealt{1992ARA&A..30..575C}; where $S \propto \nu^{-\alpha}$), and find that the K-corrected 1.4 GHz radio luminosities of two of the sources are above $10^{24}$ W Hz$^{-1}$ and thus consistent with an AGN origin (above $L_\mathrm{1.4 GHz} = 10^{24}$ W Hz$^{-1}$ the radio emission cannot be explained even by extreme star formation; e.g., \citealt{1992ARA&A..30..575C}; \citealt{2009ApJ...696..891H}; \citealt{2011ApJ...737...67M}; \citealt{2012ApJS..202....6G}; see also \citealt{2014ApJ...784..137K}). 

\begin{table*}
\begin{minipage}{\textwidth}
\centering
\caption{Radio properties of the three dwarf galaxies detected with the VLA at 3 GHz}
\label{tab:radio}
\begin{tabular}{cccccccc}
\hline
\hline 
  ID & 	$z$ 	& 	total 		& 	peak 			& Resolved &  Separation 	&$L_\mathrm{AGN}$ (1.4 GHz)	&	$L_\mathrm{bol}/L_\mathrm{Edd}$ \\
      & 		&   ($\mu$Jy)	& ($\mu$Jy beam$^{-1}$)	&		  & 	(arcsec)		&  (erg s$^{-1}$ cm$^{-2}$) 	&								\\
\hline
cid\_887 & 0.210 &	21.7$\pm$2.5	&	21.7		 	& 	no	  &	0.28			& (7.0$\pm$0.8) $\times$ 10$^{40}$	&	7 $\times$ 10$^{-2}$		\\
lid\_322 &	0.215 &	18.0$\pm$2.5	&	18.0			&	no	  &	0.35			& (6.1$\pm$0.9) $\times$ 10$^{40}$	&	8 $\times$ 10$^{-1}$		\\
lid\_325 & 0.029 &	17.2$\pm$2.5	&	17.2			&	no	  &	0.91			& (8.5$\pm1.2)$ $\times$ 10$^{38}$	&	7 $\times$ 10$^{-4}$		\\
\hline
\hline
\end{tabular}
\end{minipage}
\raggedright
\smallskip\newline\small {\bf Column designation:}~(1) \textit{Chandra} COSMOS-Legacy X-ray ID, (2) redshift, (3) total flux at 3 GHz, (4) peak flux at 3 GHz, (5) radio emission resolved, (6) separation between the radio and X-ray counterpart, (7) 1.4 GHz AGN radio luminosity derived assuming a radio spectral index $\alpha = 0.8$ and after removing the contribution from star formation, (7) Eddington ratio.
\end{table*}

To confirm the AGN origin of the radio emission, we estimate the contribution of star formation to the radio emission using the correlation between 1.4 GHz radio luminosity and $SFR$ for stars massive enough ($> 5$ M$_{\odot}$) to form supernovae (\citealt{1992ARA&A..30..575C}): 
\begin{equation}
L_\mathrm{1.4 GHz}^\mathrm{SF}  \mathrm{(W\ Hz}^{-1}\mathrm{)} = 4 \times 10^{21}\ SFR\ \mathrm{(M}_{\odot}\ \mathrm{yr}^{-1}\mathrm{)} 
\end{equation}
We find that the fraction of star formation that contributes to the $L_\mathrm{1.4 GHz}$ of the three dwarf galaxies is not significant ($\leq$ 0.02\%) and thus that all the radio emission must come from the AGN. The 1.4 GHz AGN luminosities for the three radio detected dwarf galaxies ranges from 8.49 $\times\ 10^{38}$ erg s$^{-1}$ to 6.95 $\times\ 10^{40}$ erg s$^{-1}$.

AGN accreting a low Eddington ratios are typically found to be radio-loud (e.g., \citealt{2005Ap&SS.300..219H}; \citealt{2014ApJ...787...62M}) while (low-mass) AGN with high accretion rates tend to be radio-quiet (e.g., \citealt{2006ApJ...636...56G}), in analogy with XRBs transitioning from a low/hard X-ray state in which a radio jet is present to a high/soft X-ray state in which the jet is quenched (e.g., \citealt{2006ARA&A..44...49R}). The low fraction of radio detections among the sample of 40 AGN dwarf galaxies is thus not unexpected given that most sources (95\%) are accreting at high Eddington rates, but is in agreement with the results of the scarce previous studies of jet radio emission in AGN low-mass galaxies: e.g., \cite{2006ApJ...636...56G} find only one radio-loud AGN out of a sample of 19 low-mass active galaxies, and only four other AGN dwarf galaxies have been reported to have a radio jet (NGC\,4395, \citealt{2006ApJ...646L..95W}; NGC\,404, \citealt{2012ApJ...753..103N}; Mrk\,709, \citealt{2014ApJ...787L..30R}; Henize 2-10, \citealt{2011Natur.470...66R}, \citealt{2012ApJ...750L..24R}, though the presence of an AGN in this dwarf galaxy is controversial, \citealt{2017A&A...604A.101C}). The presence of collimated radio emission is yet also possible in BHs with high accretion rates (e.g., \citealt{2010Natur.466..209P}; \citealt{2014Sci...343.1330S}; \citealt{2017ARA&A..55..303K}), which could be the case of lid\_322. The large uncertainties on the Eddington ratio could also explain the high accretion rate of this source.

The detection of jet radio emission spatially coincident with hard X-ray emission has been often used to estimate the BH mass by means of the fundamental plane of accreting BHs (e.g., \citealt{2011Natur.470...66R,2014ApJ...787L..30R}; \citealt{2011AN....332..379M};  \citealt{2012Sci...337..554W}; \citealt{2012MNRAS.424..224H}; \citealt{2017A&A...601A..20K}; \citealt{2013MNRAS.436.1546M,2013MNRAS.436.3128M,2015MNRAS.448.1893M,2018MNRAS.474.1342M}), which is a correlation between nuclear X-ray luminosity, core radio luminosity, and BH mass valid from stellar-mass BHs to SMBHs in the hard X-ray spectral state (e.g., \citealt{2004A&A...414..895F}). 
Several correlations with different scatters exist based on different samples with varied properties: e.g., the correlation from \cite{2003MNRAS.345.1057M} has the largest scatter (0.88 dex) as it includes both flat and steep radio sources, different accretion rates, and BH masses estimated using different methods; \cite{2009ApJ...706..404G} use only dynamical BH masses and nuclear radio sources, which reduces the scatter to 0.77 dex; \cite{2012MNRAS.419..267P} include only sub-Eddington accreting sources and a Bayesian approach, which yields a scatter of 0.07 dex (see e.g., \citealt{2018MNRAS.474.1342M} for a brief review). Given the range of accretion rates of the three dwarf galaxies with radio counterparts and their inferred IMBH nature (Section~\ref{BHmass}), we obtain a second estimate of their BH mass using the fundamental plane from \cite{2009ApJ...706..404G}, which is the only one proven to be valid in the IMBH regime (\citealt{2014ApJ...788L..22G}):
\begin{equation}
\begin{split}
\mathrm{log} L_\mathrm{R} = (4.80 \pm 0.24) + (0.78 \pm 0.27) \mathrm{log} M_\mathrm{BH} + \\ (0.67 \pm 0.12) \mathrm{log} L_\mathrm{X}
\end{split}
\end{equation}
where $L_\mathrm{R}$ and $L_\mathrm{X}$ are the 5 GHz core radio luminosity and 2-10 keV band nuclear X-ray luminosity, respectively, in erg s$^{-1}$. We derive $L_\mathrm{R}$ by taking the peak of 3 GHz radio emission and converting it to 5 GHz assuming $\alpha = 0.8$. For the $L_\mathrm{X}$ we take the 2-10 keV AGN X-ray luminosities derived in Section~\ref{nuclear}. 
Given the resolution of the VLA-COSMOS observations, the 5 GHz AGN radio luminosities are most likely a combination of core and lobe radio emission, hence the BH masses estimated using the fundamental plane should be taken as upper limits. We find that the BH masses are log $M_\mathrm{BH} < 6.5$ with a scatter of 0.77 dex. Further uncertainties on these mass estimates also arise from the finding of a second track in the fundamental plane (\citealt{2012MNRAS.423..590G}) and the lack of a correlation between the BH mass and the radio/X-ray plane for XRBs (\citealt{2014MNRAS.445..290G}).

\subsection{Mid-IR AGN selection}
The IR emission of reprocessed light by heated dust produces different mid-IR colors depending on whether AGN, stars or non-active galaxies are the source of the heating. Mid-IR color-color diagrams have thus become a widely used tool for identifying AGN based on either the \textit{Spitzer}/IRAC (e.g., \citealt{2004ApJS..154..166L,2007AJ....133..186L}; \citealt{2005ApJ...631..163S}) or the WISE (e.g., \citealt{2012ApJ...753...30S}) bands at 3.6$\mu$m, 4.5$\mu$m, 5.8$\mu$m, and 8.0$\mu$m. The mid-IR emission from AGN-heated dust is, in addition, less sensitive to obscuration than in the X-rays; hence mid-IR selection techniques are able to identify heavily obscured AGN missed even by deep X-ray surveys (e.g., \citealt{2008ApJ...687..111D}).

\begin{figure}
\includegraphics[width=0.48\textwidth]{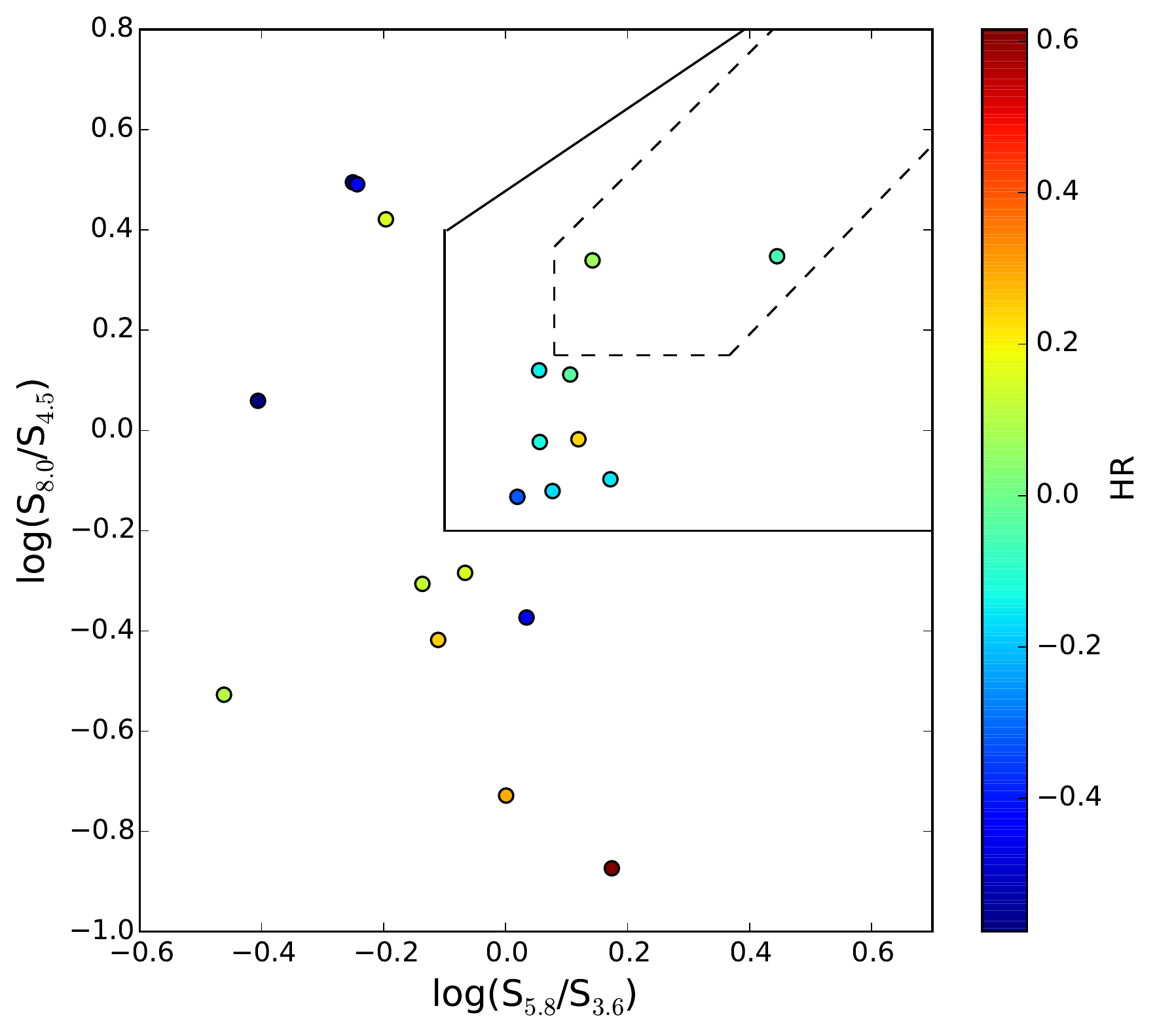}
\protect\caption[IRACplot]{IRAC color-color plot for the 20 AGN dwarf galaxies for which IRAC fluxes in the 3.6$\mu$m, 4.5$\mu$m, 5.8$\mu$m, and 8.0$\mu$m bands are available. The solid line delimites the AGN selection region of \cite{2004ApJS..154..166L,2007AJ....133..186L}. The dashed wedge is the most stringent AGN selection criteria of \cite{2012ApJ...748..142D}, which minimizes contamination from high-redshift star-forming galaxies. The $HR$ is shown as a color bar.}
\label{IRACplot}
\end{figure}

Several studies have used mid-IR color cuts for selecting AGN in dwarf galaxies (e.g., \citealt{2014ApJ...784..113S}; \citealt{2015MNRAS.454.3722S}; \citealt{2017A&A...602A..28M}). However, star-forming dwarf galaxies can have mid-IR colors similar to those of luminous AGN (\citealt{2016ApJ...832..119H}). Mid-IR color diagrams are thus not a very reliable tool for selecting AGN in dwarf galaxies. To probe this, we perform a mid-IR color-color cut for those AGN dwarf galaxies for which the four IRAC bands are available (20 out of the 40 dwarf galaxies; see Table~\ref{IRACfluxes}). Nearly half (9 out of 20) of the AGN dwarf galaxies are classified as AGN according to the IRAC criteria of \cite{2004ApJS..154..166L,2007AJ....133..186L}. When including the more stringent criteria of \cite{2012ApJ...748..142D}, which aims at minimizing contamination from high-redshift star-forming galaxies in deep IRAC surveys, only 2 of the galaxies qualify as AGN (see Fig.~\ref{IRACplot}). None of the dwarf galaxies qualify as AGN when using the \cite{2005ApJ...631..163S} magnitude color cut. These results are independent of the level of obscuration, as sources with low HR (i.e., low obscuration) are located both inside and outside the AGN locus (see color bar in Fig.~\ref{IRACplot}). 
When the AGN dominates over the galaxy light, IRAC AGN-selection is very powerful in identifying luminous both obscured and unobscured AGN (i.e., it is able to identify up to 95\% of luminous AGNs regardless of obscuration; \citealt{2010ApJ...724L..59H,2011ApJ...733..108H}). However, it is not effective in identifying low-luminosity AGN with host-dominated mid-IR SEDs nor luminous heavily obscured AGN with bright host galaxies (\citealt{2012ApJ...748..142D}). 
The UV/optical-to-IRAC bands are dominated by host galaxy light in our dwarf galaxies, most of which (90\%) are low-luminosity AGN\footnote{Low-luminosity AGN are typically defined as having $L_\mathrm{bol} \leq 10^{42}$ erg s$^{-1}$ (e.g., \citealt{2008ARA&A..46..475H}; \citealt{2014ApJ...787...62M}). We consider here the AGN X-ray luminosity instead of the bolometric luminosity just to use the same criterion as \cite{2012ApJ...748..142D}, relevant to this discussion.} with $L_\mathrm{0.5-10 keV} < 3.1 \times 10^{43}$ erg s$^{-1}$  and the rest luminous AGN ($L_\mathrm{0.5-10 keV} \geq 3.1 \times 10^{43}$  erg s$^{-1}$) with moderate to high obscuration ($HR > 0$; see Fig.~\ref{hardness}). This explains the low number of IRAC color-selected AGN dwarf galaxies, which is in agreement with the results from \cite{2012ApJ...748..142D} in which many X-ray detected AGN are outside the Lacy et al. and Stern et al. wedge. The results thus favor the use of X-ray detections as a very good method to find AGN missed by other selection criteria (i.e., mid-IR).

\begin{table}
\centering
\caption{IRAC fluxes (in $\mu$Jy)}
\label{IRACfluxes}
\begin{tabular}{ccccc}
\hline
\hline 
  ID & 	F(3.6$\mu$m) &  F(4.5$\mu$m) & F(5.8$\mu$m) & F(8.0$\mu$m)\\
\hline
 cid\_1548  &    15.1$\pm$0.1  &  18.5$\pm$0.1  &   22.4$\pm$3.1  &    14.8$\pm$4.4  \\
  cid\_232  &    64.5$\pm$0.6  &  71.6$\pm$0.4  &   73.2$\pm$2.2  &    94.4$\pm$2.7  \\
  cid\_563  &   17.9$\pm$0.2  &  19.0$\pm$0.2  &   10.1$\pm$3.8  &    59.5$\pm$3.4  \\
  cid\_887  &   26.1$\pm$0.2  &  22.4$\pm$0.1  &   19.4$\pm$4.1  &     1.2$\pm$4.4  \\
 lid\_1204  &   57.2$\pm$0.3  &  64.3$\pm$0.4  &   72.9$\pm$3.0  &     83.2$\pm$5.4  \\
 lid\_1274  &    3.86$\pm$0.08  &   5.8$\pm$0.1  &    5.8$\pm$1.6  &     0.8$\pm$1.8  \\
 lid\_1498  &   29.1$\pm$0.7  &  32.2$\pm$0.5  &   30.4$\pm$1.7  &     23.7$\pm$3.4  \\
 lid\_1755  &   92.9$\pm$0.3  &  75.1$\pm$0.5  &   59.2$\pm$6.7  &  198.1$\pm$18.1  \\
 lid\_1815  &    8.0$\pm$0.1  &   9.4$\pm$0.1  &   22.4$\pm$4.7  &    20.9$\pm$3.8  \\
 lid\_2011  &   21.7$\pm$0.1  &  23.3$\pm$0.1  &   18.6$\pm$5.1  &    12.1$\pm$4.3  \\
 lid\_2610  &     8.1$\pm$0.1  &   7.83$\pm$0.09  &    3.2$\pm$1.9  &     8.9$\pm$3.9  \\
 lid\_3076  &   15.1$\pm$0.2  &  18.7$\pm$0.2  &   16.3$\pm$3.1  &      7.9$\pm$3.2  \\
  lid\_322  &   12.4$\pm$0.1  &  15.3$\pm$0.1  &   14.8$\pm$2.1  &    11.6$\pm$4.2  \\
  lid\_325  &   131.6$\pm$0.6  &  93.6$\pm$0.7  &   45.5$\pm$5.5  &    27.8$\pm$5.8  \\
  lid\_375  &  133.5$\pm$0.6  &  96.3$\pm$0.5  &   76.2$\pm$8.3  &   298.4$\pm$14.6  \\
 lid\_3787  &    55.1$\pm$3.8  &  55.4$\pm$2.9  &  42.7$\pm$10.6  &   21.2$\pm$10.2  \\
  lid\_391  &   23.8$\pm$0.3  &  27.9$\pm$0.3  &   33.0$\pm$2.7  &    61.0$\pm$5.5  \\
 lid\_4033  &   32.0$\pm$0.2  &  26.4$\pm$0.2  &   23.4$\pm$4.7  &    13.0$\pm$5.7  \\
 lid\_4604  &   21.0$\pm$0.2  &   21.0$\pm$0.2  &   21.1$\pm$4.9  &     3.9$\pm$2.9  \\
  lid\_723  &    5.93$\pm$0.06  &   6.44$\pm$0.07  &    7.8$\pm$1.5  &     6.2$\pm$2.2  \\
  lid\_757  &    13.5$\pm$0.1  &   16.3$\pm$0.1  &   15.3$\pm$2.1  &    15.5$\pm$3.7  \\
\hline
\hline
\end{tabular}
\end{table}

\subsection{AGN fraction in dwarf galaxies}
\label{AGNfraction}
\begin{figure}
\includegraphics[width=0.48\textwidth]{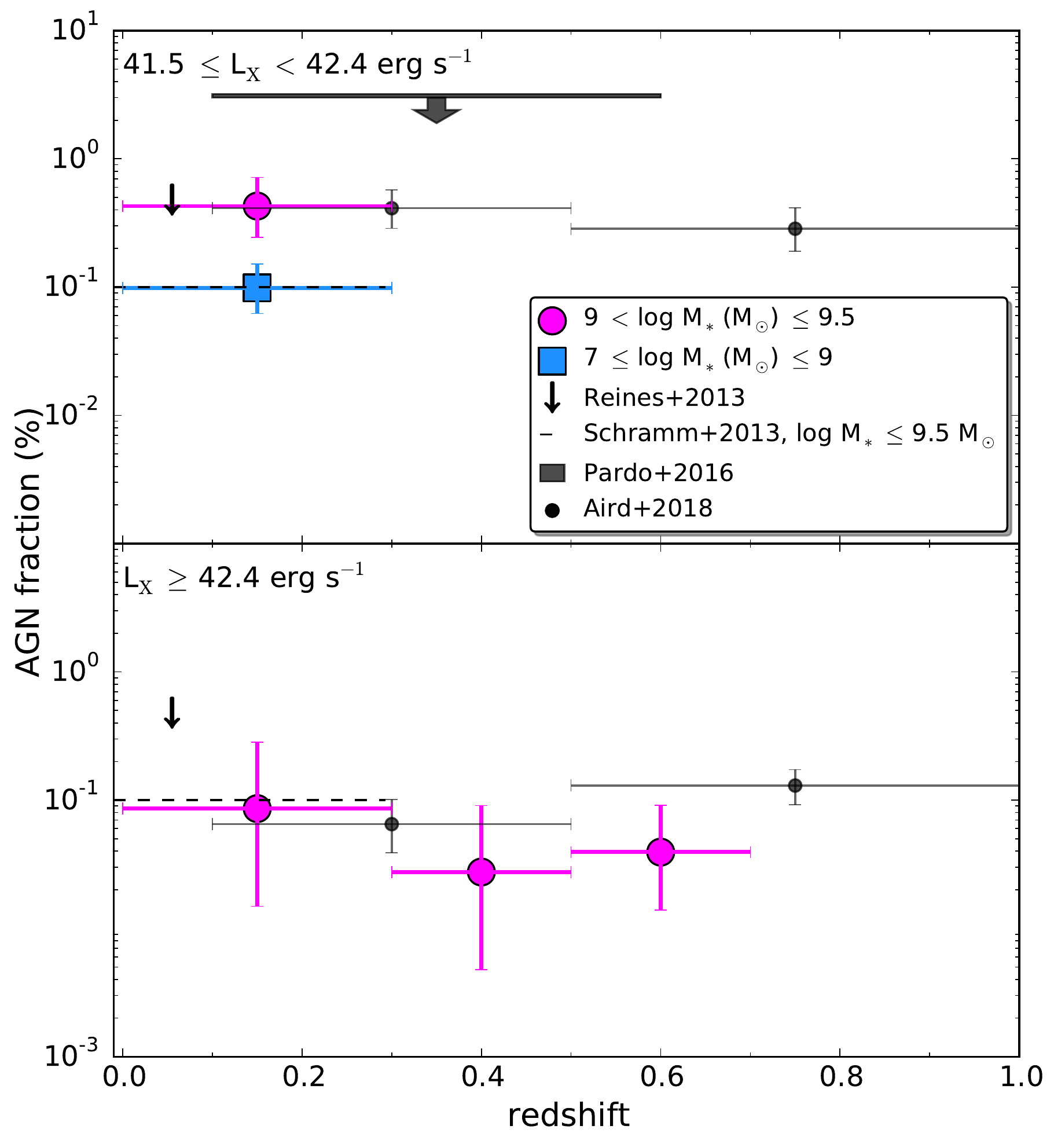}
\protect\caption[zbins]{AGN fraction as function of redshift for $3.7 \times 10^{41} \leq L_\mathrm{0.5-10 keV} < 2.4 \times 10^{42}$ erg s$^{-1}$ (top panel) and $L_\mathrm{0.5-10 keV} \geq 2.4 \times 10^{42}$ erg s$^{-1}$ (bottom panel). We further bin in stellar mass:  $10^{7}\leq M_{*}\leq10^{9}$ M$_{\odot}$ (blue square) and $10^{9}< M_{*}\leq3 \times 10^{9}$ M$_{\odot}$ (magenta circles). We plot for comparison the AGN fraction from \cite{2013ApJ...773..150S} for $M_{*} \leq 3 \times 10^{9}$ M$_{\odot}$ uncorrected from incompleteness; the upper limit from \cite{2013ApJ...775..116R} for optically-selected dwarf galaxies with $10^{7}\leq M_{*}\leq3 \times 10^{9}$ M$_{\odot}$ at $z < 0.055$, the AGN fraction from \cite{2016ApJ...831..203P} for $10^{9}\leq M_{*}\leq3 \times 10^{9}$ M$_{\odot}$ and $0.1 < z < 0.6$ (shown as an upper limit given their low statistics), and the AGN fraction from \cite{2018MNRAS.474.1225A} for 9 $<$ log M$_{*}$ (M$_{\odot}$) $<$ 9.5.}
\label{zbins}
\end{figure}

Those dwarf galaxies (52508 sources) within the area covered by the \textit{Chandra} COSMOS-Legacy survey with no X-ray emission were studied by \cite{2016ApJ...817...20M}, who divided the sources in five redshift bins out to $z$ = 1.5 ($0 < z \leq 0.3$, $0.3 < z \leq 0.5$, $0.5 <  z \leq 0.7$, $0.7 < z \leq 1$, and $1 < z \leq 1.5$) complete above the optical limit provided by the $K$-band sensitivity limit of the COSMOS survey (\citealt{2012A&A...544A.156M}; \citealt{2016ApJS..224...24L}), for which in \cite{2016ApJ...817...20M} we took a conservative value of 23.0 mag. By including the 40 AGN dwarf galaxies found here we are able to study, for the first time, the evolution with redshift, stellar mass, and X-ray luminosity of the AGN fraction in dwarf galaxies out to $z = 0.7$, down to $L_\mathrm{0.5-10 keV}\sim 10^{41}$ erg s$^{-1}$ and down to a stellar mass range $10^{7}\leq M_{*}\leq1 \times 10^{9}$ M$_{\odot}$ (Fig.~\ref{zbins}). 

We use two X-ray luminosity bins ($3.7 \times 10^{41} \leq L_\mathrm{0.5-10 keV} < 2.4 \times 10^{42}$ erg s$^{-1}$ and $L_\mathrm{0.5-10 keV} \geq 2.4 \times 10^{42}$ erg s$^{-1}$) complete above the 20\% flux limit in the 0.5-10 keV band (see Fig.~\ref{Lxfull}). The bins are determined by drawing an horizontal line from the solid curve to the left in Fig.~\ref{Lxfull}, as performed in \cite{2016ApJ...817...20M} for the stacking analysis (see Figure 2 in \citealt{2016ApJ...817...20M}), so that for all sources in each bin we are complete at 80\%. To investigate any possible loss of sources due to completeness, we also use the prescription of \citeauthor{2012ApJ...746...90A} (2012; see their sect. 5.1 for a detailed description) to compute a conditional probability density function $p(L_\mathrm{X} | M_{*},z)$ that describes the probability of a galaxy of a given $M_{*}$ and $z$ to host an AGN with luminosity $L_\mathrm{X}$. As we will see, the results obtained using both methods are consistent.

For the highest $L_\mathrm{0.5-10 keV}$ bin, we are able to derive the AGN fraction in three complete redshift bins ($0 < z \leq 0.3$, $0.3 < z \leq 0.5$, and $0.5 <  z \leq 0.7$; Fig.~\ref{zbins}, bottom panel). For the lowest $L_\mathrm{0.5-10 keV}$ bin, we are able to compute the AGN fraction in two complete stellar mass bins ($10^{7}\leq M_{*}\leq10^{9}$ M$_{\odot}$ and $10^{9}< M_{*}\leq3 \times 10^{9}$ M$_{\odot}$) for $z \leq 0.3$ (Fig.~\ref{zbins}, top panel). In each complete redshift/luminosity/stellar mass bin, the AGN fraction is calculated as $f_\mathrm{AGN}$ = $N_\mathrm{AGN}/N_\mathrm{total}$ where $N_\mathrm{total} = (N_\mathrm{undetected} + N_\mathrm{AGN})$, $N_\mathrm{AGN}$ is the number of AGN dwarf galaxies and $N_\mathrm{undetected}$ the number of dwarf galaxies with no X-ray detections from \cite{2016ApJ...817...20M} in each redshift and stellar mass bin. Because of the low number of detections, we use Poisson statistics to compute a lower and upper value of the AGN fraction in each bin complete in X-ray luminosity, redshift and stellar mass.

We note that by performing an X-ray stacking of the X-ray undetected dwarf galaxies, \cite{2016ApJ...817...20M} found an X-ray excess attributed to AGN emission. The AGN fractions derived here represent thus a lower limit on the fraction of active BHs in dwarf galaxies. Furthermore, all the AGN in the dwarf galaxies studied in this paper are of type 2. The catalog of X-ray point sources from the \textit{Chandra} COSMOS-Legacy survey (\citealt{2016ApJ...819...62C}) contains 2716 type 2 AGN and 985 type 1 AGN (\citealt{2016ApJ...817...34M}; \citealt{2017ApJ...841..102S}; H. Suh private communication). If the same ratio of type 1/type 2 AGN holds in dwarf galaxies, the addition of the type 1 sources would further increase the AGN fraction in dwarf galaxies. 

We find that the AGN fraction for $z \leq$ 0.3 is 0.43\% for $3.7 \times 10^{41} \leq L_\mathrm{0.5-10 keV} < 2.4 \times 10^{42}$ erg s$^{-1}$ and $10^{9}< M_{*}\leq3 \times 10^{9}$ M$_{\odot}$ and that it decreases with stellar mass to $f_\mathrm{AGN}$ =  0.1\% for $10^{7}\leq M_{*}\leq10^{9}$ M$_{\odot}$ (see Fig.~\ref{zbins}, top panel) and with X-ray luminosity to $f_\mathrm{AGN}$ =  0.09\% for $L_\mathrm{0.5-10 keV} \geq 2.4 \times 10^{42}$ erg s$^{-1}$ (Fig.~\ref{zbins}, top and bottom panels). The conditional probability density function for $z \leq$ 0.3 presents the same behaviour: it is $p$ = 0.46\% for $3.7 \times 10^{41} \leq L_\mathrm{0.5-10 keV} < 2.4 \times 10^{42}$ erg s$^{-1}$ and $10^{9}< M_{*}\leq3 \times 10^{9}$ M$_{\odot}$, decreases to $p$ = 0.11\% for $10^{7}\leq M_{*}\leq10^{9}$ M$_{\odot}$ and to $p$ = 0.06 \% for $L_\mathrm{0.5-10 keV} \geq 2.4 \times 10^{42}$ erg s$^{-1}$, which are in agreement with the values of the AGN fractions. 

For $z \leq$ 0.3, $10^{9}< M_{*}\leq3 \times 10^{9}$ M$_{\odot}$, and $3.7 \times 10^{41} \leq L_\mathrm{0.5-10 keV} < 2.4 \times 10^{42}$ erg s$^{-1}$, the upper value of the AGN fraction derived from Poisson statistics ($f_\mathrm{AGN}$ = 0.72\%) is consistent with the AGN fraction range of 0.6\%-3\% found by \cite{2016ApJ...831..203P} for $L_\mathrm{X} \geq 10^{41}$ erg s$^{-1}$ and $0.1 < z < 0.6$ using a sample of dwarf galaxies from the NEWFIRM Medium Band Survey with data in the AEGIS field. We note that the range of values derived by \cite{2016ApJ...831..203P} are based on very few detections (e.g., on a single detection in the case of their lower limit of 0.6\%) and therefore their AGN fraction should be considered as an upper limit given the low statistics (Fig.~\ref{zbins}, top panel). The $z \leq$ 0.3 AGN fraction for $10^{9}< M_{*}\leq3 \times 10^{9}$ M$_{\odot}$ and $3.7 \times 10^{41} \leq L_\mathrm{0.5-10 keV} < 2.4 \times 10^{42}$ erg s$^{-1}$ is also in agreement with the local AGN fraction found by \cite{2013ApJ...775..116R} ($\sim$0.5\%) for a sample of optically-selected dwarf galaxies in the same stellar mass range but not corrected for incompleteness (and thus considered as an upper limit). The AGN fraction at $z \leq$ 0.3 agrees as well with the value of $\sim$0.1\% derived by \cite{2016ApJ...831..203P} for the sample of X-ray detected dwarf galaxies found by \cite{2013ApJ...773..150S} in the \textit{Chandra} Deep Field-South, albeit of the incompleteness of the \cite{2013ApJ...773..150S} sample. \cite{2015ApJ...799...98M} find an AGN fraction for local early-type galaxies with $M_{*} < 10^{10}$ M$_{\odot}$ of $> 20\%$, which is much larger than the values found here. However, the galaxies in \cite{2015ApJ...799...98M} are highly sub-Eddington while most of the AGN dwarf galaxies found here have Eddington ratios $\gtrsim 10\%$ (as those in \citealt{2013ApJ...775..116R} and \citealt{2016ApJ...831..203P}) and are spiral and starburst.

\cite{2008ApJ...688..794S} study the evolution of AGN fraction with stellar mass, finding that for low-mass galaxies with $10^{9.7} < M_{*} < 10^{10.3}$ M$_{\odot}$ and L$_\mathrm{2-8 keV} > 10^{42}$ erg s$^{-1}$ the AGN fraction is 0.3\% for $0.1 < z < 0.4$ (see Figure 7 in their paper). Despite the slightly different stellar mass regime probed in their study, their value is in remarkably excellent agreement with the upper value of the AGN fraction ($f_\mathrm{AGN}$ = 0.28\%) at $z \leq$ 0.3 for $L_\mathrm{0.5-10 keV} \geq 2.4 \times 10^{42}$ erg s$^{-1}$. \cite{2008ApJ...688..794S} also find that the AGN fraction increases with stellar mass, e.g., from 0.3\% for $M_{*} < 10^{10.3}$ to 1\% for $M_{*} > 10^{10.5}$ M$_{\odot}$ and $0.1 < z < 0.4$, in agreement with the increase of AGN fraction with stellar mass for galaxies up to $10^{11}$ M$_{\odot}$ found by \cite{2003MNRAS.346.1055K} for $z < 0.3$. Although we probe a much narrower range of stellar masses, such a trend is also observed in Fig.~\ref{zbins} (top panel), in which the AGN fraction for $z \leq 0.3$ increases from 0.1\% for $10^{7}\leq M_{*}\leq10^{9}$ M$_{\odot}$ to 0.43\% for $10^{9}< M_{*}\leq3 \times 10^{9}$ M$_{\odot}$.  

\cite{2012ApJ...746...90A,2018MNRAS.474.1225A} have studied in detail the evolution of AGN fraction with redshift, X-ray luminosity and stellar mass using data from the PRIMUS survey, COSMOS, and the CANDELS fields, probing stellar masses down to 8.5 $<$ log $M_{*} <$ 9 M$_{\odot}$ and X-ray luminosities down to $L_\mathrm{2-10 keV} \sim 10^{41}$ erg s$^{-1}$. The authors find that the AGN fraction decreases with decreasing stellar mass, in agreement with our results and with previous studies, and that for log $M_{*}$ =  9.25 M$_{\odot}$ and log $L_\mathrm{0.5-10 keV} \sim$ 41.4 erg s$^{-1}$ (J. Aird, private communication) the AGN fraction seems to drop with redshift (albeit with large uncertainties, so it could as well be consistent with constant). To compare these results to ours, we plot in Fig.~\ref{zbins} (bottom panel) the AGN fraction found by \cite{2018MNRAS.474.1225A} for star-forming galaxies with 9 $<$ log $M_{*}$ (M$_{\odot}$) $<$ 9.5 at redshift $z\sim$0.3 and $z\sim$ 0.7. At $z\sim$0.3, their values are fully consistent with ours both for $3.7 \times 10^{41} \leq L_\mathrm{0.5-10 keV} < 2.4 \times 10^{42}$ erg s$^{-1}$ and $L_\mathrm{0.5-10 keV} \geq 2.4 \times 10^{42}$ erg s$^{-1}$. 

For $L_\mathrm{0.5-10 keV} \geq 2.4 \times 10^{42}$ erg s$^{-1}$, we also find a possible decrease of AGN fraction with redshift, at least out to $z$ = 0.5, of the form $f_\mathrm{AGN} \propto z^{-0.3 \pm 0.1}$ (correlation coefficient $r^2$ = 0.6). We note though that the values are consistent when considering the uncertainties derived from Poisson statistics. The behavior of the AGN fraction of dwarf galaxies is, in any case, significantly different than that of more massive galaxies (for which there is a significant increase in AGN fraction from $z\sim$0.1 to $z\sim$2; see Figure 6 in \citealt{2018MNRAS.474.1225A}), suggesting that the AGN evolution in dwarf galaxies is very different to that of massive galaxies and that BH growth seems to be suppressed in the low-mass regime (\citealt{2018MNRAS.474.1225A})

The low AGN fraction in dwarf galaxies and their possible decrease with redshift (Fig.~\ref{zbins}, bottom panel) could be explained if the BH switches on and off on very short timescales, specially at high redshifts, in agreement with observational constraints (e.g., \citealt{2015MNRAS.451.2517S}) and models in which BH growth is dominated by short episodes of accretion at high Eddington rates (e.g., \citealt{2005ApJ...633..624V}; \citealt{2014ApJ...784L..38M}; \citealt{2015MNRAS.451.1964S}; \citealt{2016MNRAS.458.3047P}; \citealt{2017MNRAS.472L.109A}). This could also explain the low number of detections of the high-$z$ SMBH progenitors, which, even when taking obscuration into account, should be bright enough to be detected by current wide-area X-ray surveys such as \textit{Chandra} COSMOS Legacy (\citealt{2017MNRAS.466.2131P}). 

Numerical simulations also predict that the early growth of BHs in low-mass galaxies is significantly suppressed by bursty stellar feedback, which continuously evacuates gas from the nucleus, so that efficient BH growth begins when the stars dominate the gravitational potential in the nucleus and star formation becomes less bursty, which roughly corresponds to galaxies growing to $M_{*} > 10^{9.5}$ M$_{\odot}$ (e.g., \citealt{2015MNRAS.452.1502D}; \citealt{2017MNRAS.468.3935H}; \citealt{2017MNRAS.465...32B}; \citealt{2017MNRAS.472L.109A}). This dominance of supernova feedback in low-mass galaxies has been also proven observationally (\citealt{2018ApJ...855L..20M}) and could explain the low AGN fraction found in dwarfs compared to that of higher mass galaxies even when assuming a BH occupation fraction of 100\% independent of galaxy mass. 

The AGN fraction, and its dependence on stellar mass and redshift, is expected to be dependent on bolometric AGN luminosity  (e.g., see Figure 23 in \citealt{2015ApJ...811...26T}, Figure 10 in \citealt{2016MNRAS.460.2979V}). Our results showcase such a behavior for $z \leq$ 0.3, where the AGN fraction decreases with X-ray luminosity for a given stellar mass range. At low $L_\mathrm{bol}$ (i.e., $L_\mathrm{bol}$ $\rightarrow{0}$) the AGN fraction can be taken as a proxy for BH occupation fraction; however, for high $L_\mathrm{bol}$ (i.e., $L_\mathrm{bol} > 10^{41}$ erg s$^{-1}$) the AGN fraction is a lower limit to the BH occupation fraction (\citealt{2016MNRAS.460.2979V}). Even for the lowest X-ray luminosity bin probed here ($3.7 \times 10^{41} \leq L_\mathrm{0.5-10 keV} < 2.4 \times 10^{42}$ erg s$^{-1}$) and assuming a conservative bolometric correction factor $k = 5$ (i.e., in between those of AGN and stellar-mass sources; \citealt{2015MNRAS.448.1893M}), the bolometric luminosity is $L_\mathrm{bol} \geq 1.9 \times 10^{42}$ erg s$^{-1}$ and thus the AGN fraction at $z \leq$ 0.3 can only be taken as a very rough lower limit to the BH occupation fraction. This does not allow us to draw any firm conclusions on the formation mechanism of seed BHs in the early Universe, which is expected to be dominated by direct collapse if a low occupation fraction is found in local dwarf galaxies and by Population III stars if the occupation fraction in local dwarf galaxies is high (e.g., \citealt{2008MNRAS.383.1079V}; \citealt{2010A&ARv..18..279V}; \citealt{2010MNRAS.408.1139V}). Yet we note that a decrease of BH occupation fraction with stellar mass is also expected from simulations of seed BH formation (e.g., \citealt{2008MNRAS.383.1079V,2016MNRAS.460.2979V}; \citealt{2011ApJ...742...13B}; \citealt{2017MNRAS.468.3935H}). The low AGN fraction found in dwarf galaxies and its decrease with decreasing stellar mass seem to favor the scenario in which seed BHs formed from direct collapse (see also \citealt{2017IJMPD..2630021M}); however, this might as well just be an effect caused by more massive BHs being easier to detect than light BHs and by the short timescales of the AGN activity cycle. 

The future X-ray mission concept \textit{Lynx} would allow us to make a significant leap forward in the study of IMBHs, in particular at high redshift. Future large-area surveys (at least 2 deg$^{2}$) and 2-3 orders of magnitude fainter than \textit{Chandra} COSMOS Legacy would be crucial to provide samples large enough to fill the relevant parameter space (mass/luminosity/environment) but also deep enough to approach the low X-ray luminosities in which good constraints on the BH occupation fraction, and thus on the formation model of the SMBH progenitors, can be obtained. In this context, \textit{Lynx} will be able to push the current sensitivity of \textit{Chandra} COSMOS Legacy out $z=2$ and to detect $5 \times 10^{5}$ M$_{\odot}$ BHs with L$_\mathrm{X}$ = 10$^{41}-10^{42}$ erg s$^{-1}$ at $z$ = 10. \textit{Lynx} positional accuracy and spatial resolution will be also needed for these studies to securely associate multiwavelength data to the X-ray emission.

\section{Conclusions}
\label{conclusions}
The presence of seed BHs at $z > 7$ was invoked in order to explain the finding of SMBHs when the Universe was only $\sim$0.8 Gyr old. While detecting these SMBH progenitors in the early Universe constitutes an observational challenge, those seeds that did not grow into SMBHs should be observed in low-mass galaxies at lower redshifts. When actively accreting, they can be easily detected as AGN by means of optical, IR or X-ray searches. This has already yielded the detection of a few hundreds of AGN in local dwarf galaxies, most of them at $z < 0.5$. 

In this paper we report the discovery of 40 AGN located in dwarf galaxies ($10^{7}\leq M_{*}\leq3 \times 10^{9}$ M$_{\odot}$) out to $z \sim 2.4$. This constitutes the highest-redshift sample of AGN in low-mass galaxies, with 12 sources being at $z > 0.5$. 
Most of the host galaxies (39 out of 40) are star-forming; yet, their X-ray luminosity is one order magnitude higher than the typical X-ray luminosity of XRBs. The contribution from XRBs to the 0.5-10 keV X-ray emission is $\lesssim$16 \%. After removing the XRB and hot ISM contribution to the X-ray emission, the AGN luminosities range $L_\mathrm{0.5-10 keV} \sim 10^{39}-10^{44}$ erg s$^{-1}$. The brightest X-ray source is cid\_1192, which is also the highest-redshift AGN in our sample ($z = 2.39$) and the new record-holder of an AGN in a dwarf galaxy. Based on the hardness ratios, we find that 60\% of the sources are obscured and that the level of obscuration increases with redshift.

Using the recent scaling relation between BH and stellar mass of \cite{2015ApJ...813...82R}, we find a range of BH masses of $M_\mathrm{BH} \sim 10^{4} - 7 \times 10^{5}$ M$_{\odot}$ and thus that all the AGN are consistent with hosting IMBHs. The BH mass is $< 3 \times 10^{4}$ M$_{\odot}$ for four of the AGN dwarf galaxies, which makes them the lightest IMBH candidates ever found in low-mass galaxies. One of the AGN dwarf galaxies has $M_\mathrm{*} = 6.6 \times 10^{7}$ M$_{\odot}$ and constitutes the first detection of an AGN in a galaxy with $M_\mathrm{*} < 10^{8}$ M$_{\odot}$. Most of the sources (38 out of 40) have Eddington ratios $>$ 1\% and three have VLA radio emission with flux densities above 17 mJy. Future near-IR follow-up observations are planned with the aim of constraining further the BH mass and properties of some of these extreme AGN. 

Adding the new 40 AGN dwarf galaxies to the sample of X-ray undetected dwarf galaxies in the \textit{Chandra} COSMOS-Legacy survey (\citealt{2016ApJ...817...20M}), we are able to derive the AGN fraction in dwarf galaxies out to $z$ = 0.7 and, for the first time, down to a stellar mass range of $10^{7}\leq M_{*}\leq3 \times 10^{9}$ M$_{\odot}$, as well as to study with completeness its evolution with stellar mass, X-ray luminosity and redshift. For $z \leq$ 0.3, we find that the AGN fraction is 0.43\% for $10^{9}< M_{*}\leq3 \times 10^{9}$ M$_{\odot}$ and $L_\mathrm{0.5-10 keV} \sim 10^{41}-10^{42}$ erg s$^{-1}$ and that it decreases with X-ray luminosity and with decreasing stellar mass. For $L_\mathrm{0.5-10 keV} \gtrsim 10^{42}$ erg s$^{-1}$ the AGN fraction is of 0.09\% and tentatively decreases with redshift (albeit large errors). This behavior clearly differs from that observed in massive galaxies ($>10^{10}$ M$_\mathrm{*}$), suggesting that BH growth is quenched in dwarf galaxies. 

The presence in the early Universe of heavy seed BHs formed from direct collapse of pregalactic gas disk is thought to be less common than that of light seeds formed from Population III stars (e.g., \citealt{2008MNRAS.383.1079V}; \citealt{2010A&ARv..18..279V}). Although the AGN fractions reported here constitute a merely lower limit to the true BH occupation fraction in local dwarf galaxies, the low values found and their decrease with decreasing stellar mass suggest that seed BHs could have formed predominantly through direct collapse. However, this finding might as well just be a reflection of heavier BHs being easier to detect than light ones. The low values of AGN fraction, specially at high redshift, are also expected in BH growth models in which seed BHs undergo short phases of super-Eddington accretion, which could also explain the low detection rate of high-$z$ seed BHs (e.g., \citealt{2017MNRAS.466.2131P}; \citealt{2017arXiv170606592P}). As suggested by \cite{2017MNRAS.466.2131P}, wide-area X-ray surveys should be able to detect the progenitors of SMBHs at high redshift. 
We have proven that this is the case with \textit{Chandra} COSMOS-Legacy, which has allowed us to boost the detections of IMBHs from the local Universe to the epoch of the pinnacle of star formation and BH activity. Increasing the depth of this survey (e.g., to 250 ks) we would be able to detect AGN in dwarf galaxies out to $z$ = 0.5 for $L_\mathrm{X} \sim 10^{41}$ and out to $z$ = 1 for $L_\mathrm{X} > 10^{42}$, which would provide stronger constraints on BH seed formation models and allow us to better understand how SMBHs formed in the early Universe.

\section*{Acknowledgments}
The authors thank J. Aird and M. Elvis for insightful discussion. This work was supported in part by NASA \textit{Chandra} grant G05-16099X (M.M.), GO3-14150C and GO3-14150B (F.C, S.M.).


\bibliographystyle{mnras}
\bibliography{/Users/mmezcua/Documents/referencesALL}


\bsp	
\label{lastpage}
\end{document}